\documentclass[preprint2]{aastex}

\shorttitle{X-rays from massive stars in Cyg\,OB2}
\shortauthors{Rauw et al.}

\usepackage{rotating}

\begin{document}


\title{X-ray emission from massive stars in Cyg\,OB2}


\author{G. Rauw and Y. Naz\'e\altaffilmark{1}}
\affil{Institut d'Astrophysique \& G\'eophysique, Universit\'e de Li\`ege, Belgium}
\email{rauw@astro.ulg.ac.be}
\author{N.J. Wright\altaffilmark{2}, J.J. Drake, and M.G. Guarcello}
\affil{Smithsonian Astrophysical Observatory, Cambridge, USA}
\author{R.K. Prinja and L.W. Peck}
\affil{Department of Physics \& Astronomy, University College London, United Kingdom}
\author{J.F. Albacete Colombo}
\affil{Centro regional Zona Atl\'antica, Universidad Nacional del Comahue, Viedma, Argentina}
\author{A. Herrero}
\affil{Instituto de Astrof\'{\i}sica de Canarias, Universidad de La Laguna, La Laguna, Spain}
\author{H.A. Kobulnicky}
\affil{Department of Physics \& Astronomy, University of Wyoming, Laramie, USA}
\author{S. Sciortino}
\affil{INAF - Osservatorio Astronomico di Palermo, Italy}
\and
\author{J.S. Vink}
\affil{Armagh Observatory, College Hill, Armagh, Northern Ireland}
\altaffiltext{1}{Research Associate FRS-FNRS, Belgium}
\altaffiltext{2}{present address: Centre for Astrophysics Research, University of Hertfordshire, Hatfield, UK}
\begin{abstract}
We report on the analysis of the {\it Chandra}-ACIS data of O, B and WR stars in the young association Cyg\,OB2. X-ray spectra of 49 O-stars, 54 B-stars and 3 WR-stars are analyzed and for the brighter sources, the epoch dependence of the X-ray fluxes is investigated. The O-stars in Cyg\,OB2 follow a well-defined scaling relation between their X-ray and bolometric luminosities:  $\log{\frac{{\rm L}_{\rm X}}{{\rm L}_{\rm bol}}} = -7.2 \pm 0.2$. This relation is in excellent agreement with the one previously derived for the Carina OB1 association. Except for the brightest O-star binaries, there is no general X-ray overluminosity due to colliding winds in O-star binaries. Roughly half of the known B-stars in the surveyed field are detected, but they fail to display a clear relationship between L$_{\rm X}$ and L$_{\rm bol}$. Out of the three WR stars in Cyg\,OB2, probably only WR\,144 is itself responsible for the observed level of X-ray emission, at a very low $\log{\frac{{\rm L}_{\rm X}}{{\rm L}_{\rm bol}}} = -8.8 \pm 0.2$. The X-ray emission of the other two WR-stars (WR\,145 and 146) is most probably due to their O-type companion along with a moderate contribution from a wind-wind interaction zone. 
\end{abstract}





\section{Introduction}
Cygnus OB2 is not only interesting as a very active star forming region, but also as an association containing a wealth of massive stars. This population has been intensively studied over recent years both in terms of a general census \citep{Comeron, Hanson,Negueruela, Wright} as well as in terms of an extensive radial velocity survey (\citet{Kiminki}, and \citet{Kobulnicky} for the most recent results). Despite its heavy extinction, Cyg\,OB2 is therefore an interesting place for the study of massive stars over a wide range of wavelengths, including the X-ray domain.

X-ray emission is a well-known property of massive stars of spectral type earlier than about mid-B. For single O-stars, this emission is generally attributed to a distribution of hydrodynamic shocks produced by the so-called Line Deshadowing Instability \citep[LDI, e.g.][]{Feldmeier} in the radiatively-driven winds of these objects. Another mechanism to produce X-ray emission from single early-type stars is the head-on collision of magnetically channeled gas in the stellar winds of massive stars that feature a strong enough magnetic field \citep[e.g.][]{BabelMontmerle,ud-Doula}. Moreover, in massive binary systems, additional X-ray emission can arise from large-scale shocks associated with wind-wind interactions \citep[e.g.][]{SBP}. The shocks in magnetically confined winds and in colliding wind binaries occur at much higher Mach numbers than LDI shocks and are thus expected to produce stronger and harder X-ray emission than the latter. 
Already in the early-days of the discovery of X-ray emission of early-type stars with the {\it EINSTEIN} satellite \citep{Harnden}, it has been found that the X-ray luminosity of O-type stars scales with their bolometric luminosity \citep[e.g.][]{Sciortino}. This relationship was subsequently confirmed and refined with large samples of O-type stars observed with {\it ROSAT} \citep{Berghoefer}, and more recently {\it XMM-Newton} \citep{YN}. This situation contrasts with that of Wolf-Rayet stars, for which there is no clear dependence between X-ray and bolometric luminosities \citep{Wessolowski} and where some Wolf-Rayet stars remain undetected with current observatories \citep[e.g.][]{Oskinova,Gosset2}. The same holds for the lower luminosity end of massive stars, where only a subsample of the B-type stars is detected \citep[e.g.][]{Berghoefer}. In the case of non-supergiant B-stars, the winds are generally thought to be too tenuous to produce strong emission via LDI shocks and alternative scenarios such as low-mass pre-main sequence companions and magnetic wind confinement have been suggested \citep{Evans}. 

Despite some attempts for a theoretical explanation \citep{OC99}, the origin of the empirical L$_{\rm X}$/L$_{\rm bol}$ scaling relation of O-stars remained elusive for many years. From first principles, a steeper than linear relation would be expected for X-ray emission produced by LDI shocks\footnote{ L$_{\rm X}$ should scale as L$_{\rm bol}^{1.7}$ or L$_{\rm bol}^{3.4}$ respectively for radiative or adiabatic shocks \citep{Owocki}.}. Very recently, \citet{Owocki} argued that the shocks in O-star winds are radiative, although the density of the winds remains in most cases sufficiently low to prevent the wind absorption from playing a significant role. Turbulence in the radiatively cooling post-shock gas, could then lead to an efficient mixing of cold and hot material. Assuming a scaling of the volume filling factor of the hot gas with some ad-hoc power $m \simeq 0.2 - 0.4$ of the ratio between cooling length and position in the wind, \citet{Owocki} were able to recover the observed L$_{\rm X}$/L$_{\rm bol}$ scaling relation. 
Furthermore, the \citet{Owocki} scenario predicts a change in the behavior of the L$_{\rm X}$ versus L$_{\rm bol}$ relation at the high- and low-luminosity ends of the O-star domain, which needs to be tested observationally. Indeed, at the high-luminosity end, winds should become optically thick\footnote{This transition towards optically thick winds was also found by \citet{Vink2} in Monte Carlo radiative transfer models. These authors found a kink in the relation between $\dot{M}$ and the Eddington factor $\Gamma$ corresponding to the transition in spectral morphology from normal O/Of-type stars to Wolf-Rayet characteristics.}, whereas at the low-luminosity end the shocks should become adiabatic, resulting in a change of the L$_{\rm X}$ versus L$_{\rm bol}$ relation. 

\begin{figure*}[thb]
\begin{minipage}{8cm}
\begin{center}
\resizebox{8cm}{!}{\includegraphics{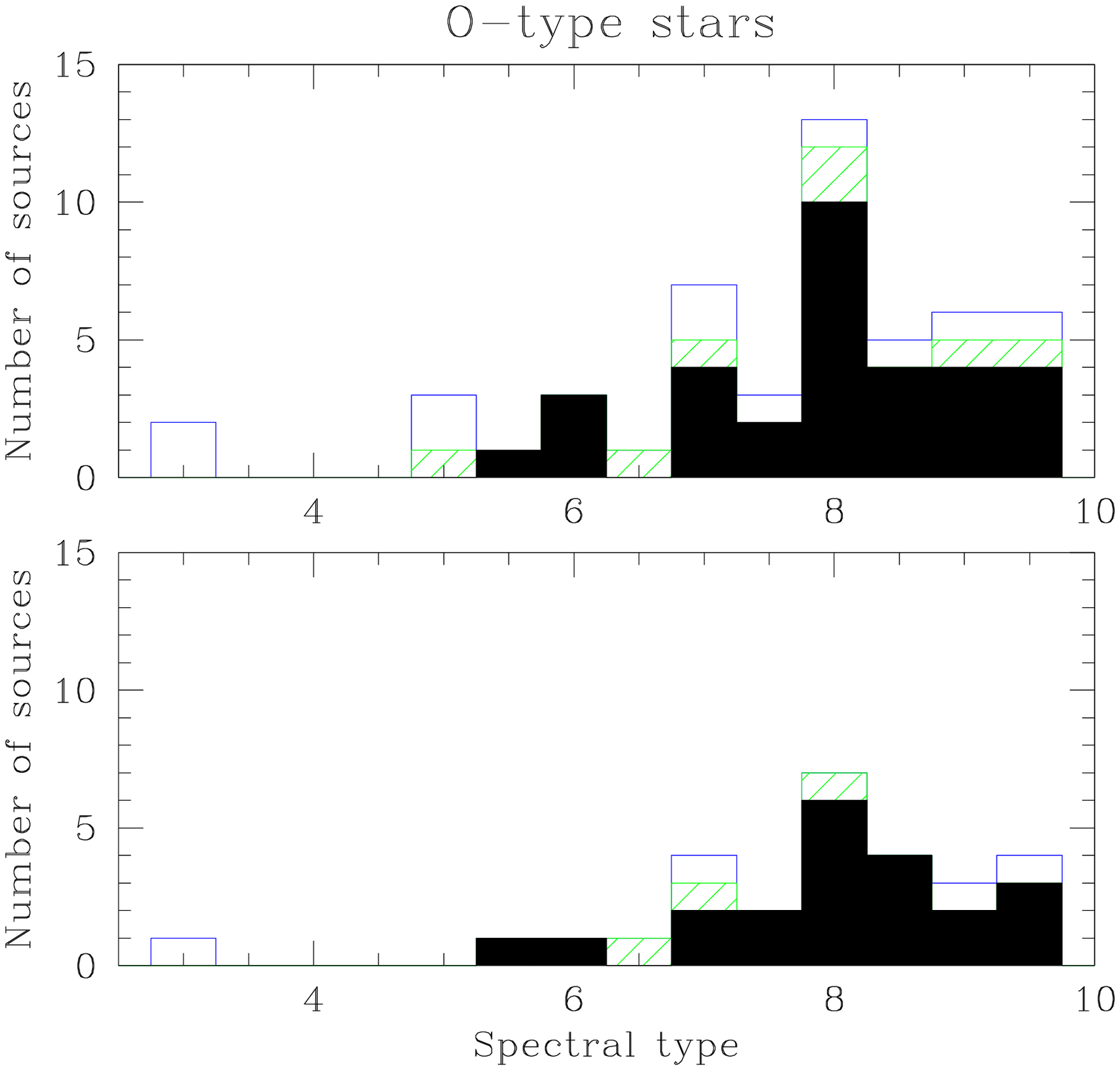}}
\end{center}
\end{minipage}
\hfill
\begin{minipage}{8cm}
\begin{center}
\resizebox{8cm}{!}{\includegraphics{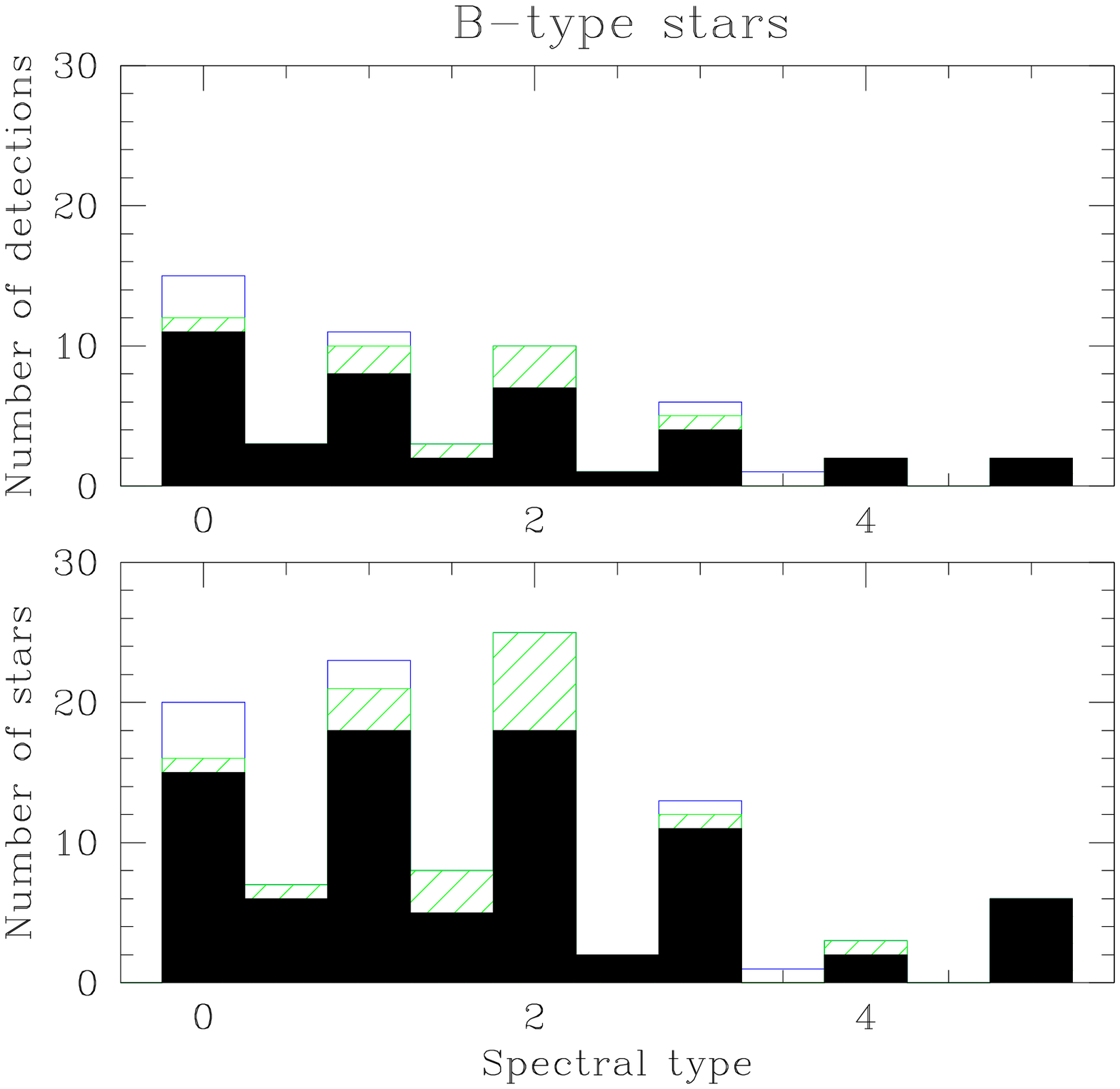}}
\end{center}
\end{minipage}
\caption{Left: histogram of the detected O-stars as a function of spectral types. The upper panel shows the distribution of the full sample, where the spectral type of the primary is used for binary systems. The lower panel illustrates the situation when only presumably single stars are considered. The filled, hatched and open histograms stand for main-sequence stars, giants and supergiants respectively. Right: histogram of the B-type stars as a function of spectral type. The upper panel illustrates the distribution of spectral types for those stars that are detected as counterparts of X-ray sources, whilst the lower panel corresponds to the full sample of known B-stars in the field of view. The various types of histograms have the same meaning as for O-stars. \label{sample}}
\end{figure*}

The {\it Chandra} Cyg\,OB2 Legacy Survey offers an ideal data set for such an in-depth study of the X-ray properties of massive stars, as it provides a large and homogeneous sample of objects from early B-type, over almost all categories of O-stars, and even several Wolf-Rayet stars.  

\section{Data analysis}
The data analyzed here are taken from the {\it Chandra} Cygnus\,OB2 Legacy Survey. A full description of this project and details on the data reduction are given by \citet{Drake} and \citet{Wright1}. The survey consists in an overlapping $6 \times 6$ raster mosaic of 30\,ks exposures with an 8\,arcmin pointing offset between adjacent fields. The sources in the central 42\,arcmin$^2$ square region were typically observed four times at different off-axis angles. The survey reaches 90\% completeness for $L_X = 7 \times 10^{29}$\,erg\,s$^{-1}$ at the distance of Cyg\,OB2. For our variability study of O-type stars, we further include some spectra from the {\it XMM-Newton} observations discussed by \citet{Rauw}, \citet{Naze9}, and \citet{Cazorla}.

\subsection{The sample of massive stars}
Cyg\,OB2 has a very rich population of massive stars. The field of view of the {\it Chandra} survey contains a hundred stars classified as  B-stars, 52 O-type stars and 3 Wolf-Rayet stars \citep{Wright}\footnote{In the present study, we adopt the spectral types compiled and homogenized by \citet{Wright}.}.
\subsubsection{O-stars}
The sample of O-stars detected with ACIS spans a wide range in spectral types and luminosity classes, from O3\,I to O9.5\,V, although there is a clear dominance of spectral type O8\,V (see Fig.\,\ref{sample}). All 52 known O-stars in the field of view are detected as X-ray emitters. The information on multiplicity of our stars was taken from the latest results of the Cygnus OB2 Radial Velocity Survey \citep[][and Kobulnicky et al., in preparation]{Kobulnicky,Kiminki}. 

We have estimated the interstellar neutral hydrogen column density towards each source from the $E(B-V)$ color excess following the conversion formula of \citet{Bohlin}. The color excess was evaluated from the observed $B-V$ and the intrinsic $(B-V)_0$ colors as a function of spectral type according to \citet{MP}. For stars, where no $B-V$ data are available, we used the $E(J-K)$ color-excess \citep{Negueruela} and the relation $E(J-K) = 0.525\,E(B-V)$ from \citet{RL}. As expected for observations in Cyg\,OB2, the interstellar absorption is quite heavy, leading to N$_{\rm H}$ values that frequently exceed $10^{22}$\,cm$^{-2}$ (see Fig.\,\ref{sample2}). 

\subsubsection{B-stars}
Out of 108 B-type stars that fall into the field of view covered by the survey, 54 are detected as X-ray sources. The information on spectral types and multiplicity is taken from \citet{Wright}. Unlike the situation for O-stars, the B-star sample probably suffers from severe incompleteness and our knowledge of its multiplicity is only fragmentary.

\begin{figure}[htb]
\begin{center}
\resizebox{8cm}{!}{\includegraphics{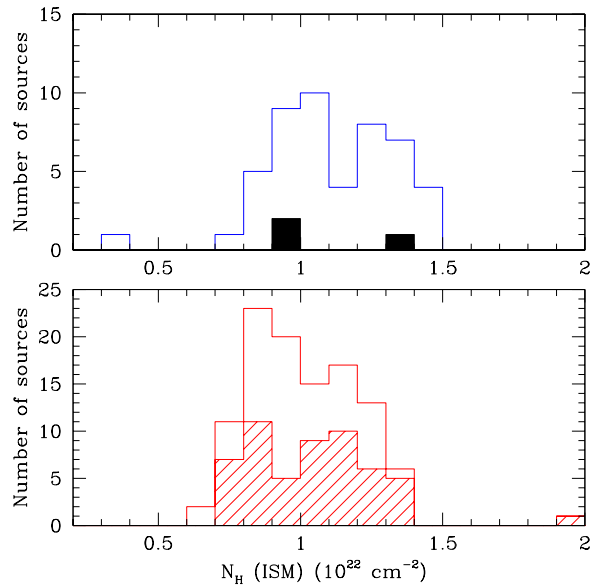}}
\end{center}
\caption{Top: histogram of the ISM neutral hydrogen column density towards the O and WR-stars in our sample. The filled histogram refers to the 3 Wolf-Rayet stars. Bottom: histogram of the ISM neutral hydrogen column density towards the B-type stars. The shaded histogram corresponds to the distribution for the detected objects. \label{sample2}}
\end{figure}

Out of the nine known binaries in the sample, all but one are detected in X-rays. As can be seen in Fig.\,\ref{sample}, B0 stars have a higher detection rate (75\%) than B2 stars (40\%). The distribution of the interstellar column density (Fig.\,\ref{sample2}) clearly indicates that absorption by the ISM has no impact on the detection or non-detection of an X-ray source associated with a B-star. 
\subsubsection{Wolf-Rayet stars}
There are three Wolf-Rayet stars that fall in the field of view of the {\it Chandra} survey: WR\,144, 145 and 146. All of them are detected as X-ray sources and they are all of the WC subclass, with WR\,146 being one of the few WN/WC hybrid stars \citep{vdH}. Two of these three Wolf-Rayet stars (WR\,145 and WR\,146) are actually part of binaries or higher multiplicity systems. All three stars have interstellar hydrogen column densities around $10^{22}$\,cm$^{-2}$ (see Fig.\,\ref{sample2}) and are considered probable members of Cyg\,OB2 \citep{LS}. 

\subsection{Spectral fitting}
The ACIS X-ray spectra were binned in such a way as to have at least 5 counts per energy bin. They were then fitted using {\tt xspec} v.12.7 \citep{Arnaud}. For each source, we fitted both the spectra from different observations, provided there were sufficient counts in the individual spectra\footnote{We require that the binned spectra must have at least four independent energy bins.}, and the total, combined ACIS spectrum. For the analysis of the global properties of the X-ray spectra (Sect.\ \ref{global}), we used only the combined ACIS spectra. 

Unless stated otherwise, we used optically thin thermal plasma \citep[{\tt apec},][]{apec} models with solar abundances according to \citet{AG}\footnote{For CCD spectra such as those investigated here, the revision of the solar composition \citep{Asplund} has little impact on the results of the fit.}. 
The majority of the spectra were fitted assuming a single plasma component. Only in those cases where the spectra are of sufficient quality and where the addition of a second plasma temperature significantly improves the fit, did we opt for a two-temperature model. The models used were thus:
{\tt phabs*phabs*apec} and {\tt phabs*phabs*(apec+apec)}, where the {\tt phabs} component indicates the absorption model. The first absorption column was fixed to the interstellar neutral hydrogen column density derived above.
The second absorption column is meant to represent the absorption by the stellar wind. At first sight, this might appear a crude approximation as the wind material is ionized by the photospheric radiation, whilst the model employed here assumes neutral material. However, this approximation impacts the spectrum only at energies below 1\,keV, which are anyway absorbed by the heavy interstellar absorption in Cyg\,OB2. Another approximation comes from the fact that in real stellar winds, the absorbing and emitting materials are interleaved and a more sophisticated treatment of absorption is needed to fit high-resolution X-ray spectra of O-type stars \citep{HRN}. Yet, for CCD spectra, such as those analyzed here, such complex models cannot be fitted and the above simple models are sufficient to provide a general description of the X-ray spectral energy distribution. 

\subsection{Pile-up}
The ACIS spectra of the four brightest sources in our sample (Cyg\,OB2 \#8a, \#5, \#9 and \#12) suffer from severe pile-up. {\it XMM-Newton} observations of these objects revealed orbital or long-term changes in the observed X-ray flux \citep[][and references therein]{Cazorla}.  

We have evaluated the pile-up fraction of the ACIS spectra via two independent techniques. First, we applied PIMMS to the best-fit parameters inferred from the {\it XMM-Newton} spectra \citep{DeBecker,Blomme,Linder,Naze9,Cazorla}. Adopting the lowest fluxes observed with {\it XMM-Newton}, we estimated pile-up fractions of 60, 34, 18 and 16\% respectively for on-axis {\it Chandra} observations of Cyg\,OB2 \#8a, \#5, \#9 and \#12. Of course, the sources were not observed on-axis for each pointing of the {\it Chandra} campaign. For large off-axis angles, the degradation of the PSF reduces pile-up. Therefore, in a second approach, we estimated the pile-up fraction directly from the data, by looking at the maximum count rate of all the pixels within the PSF. This approach confirms that the four brightest sources are most of the time subject to pile-up fractions well above 10\%. The {\it Chandra} data of these sources are thus not considered further in this paper.

\begin{figure}[t!hb]
\begin{center}
\resizebox{8cm}{!}{\includegraphics{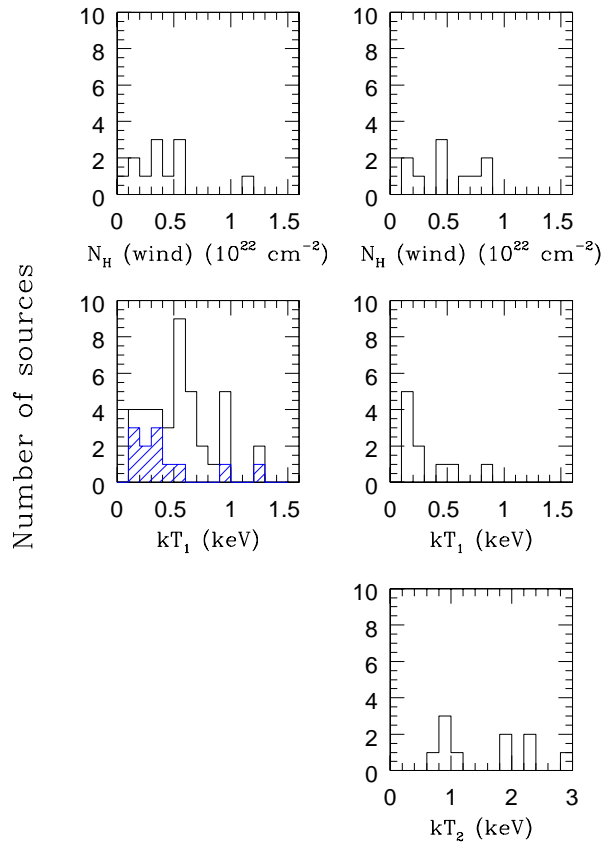}}
\end{center}
\caption{Histograms of the wind column densities and temperatures obtained from our fits of the spectra of 49 O-type stars in Cyg\,OB2. The left column corresponds to the 1-T fits (39 stars), whilst the right column yields the results for 2-T models (10 objects). 27 of the 39 objects with 1-T model fits have a zero wind column density. These objects are not shown in the N$_{\rm H}$ (wind) histogram. The hatched histogram in the distribution of $kT$ for 1-T fits corresponds to objects with a significant wind column density.\label{stats}}
\end{figure}

\begin{figure*}[h!tb]
\begin{minipage}{8cm}
\begin{center}
\resizebox{8cm}{!}{\includegraphics{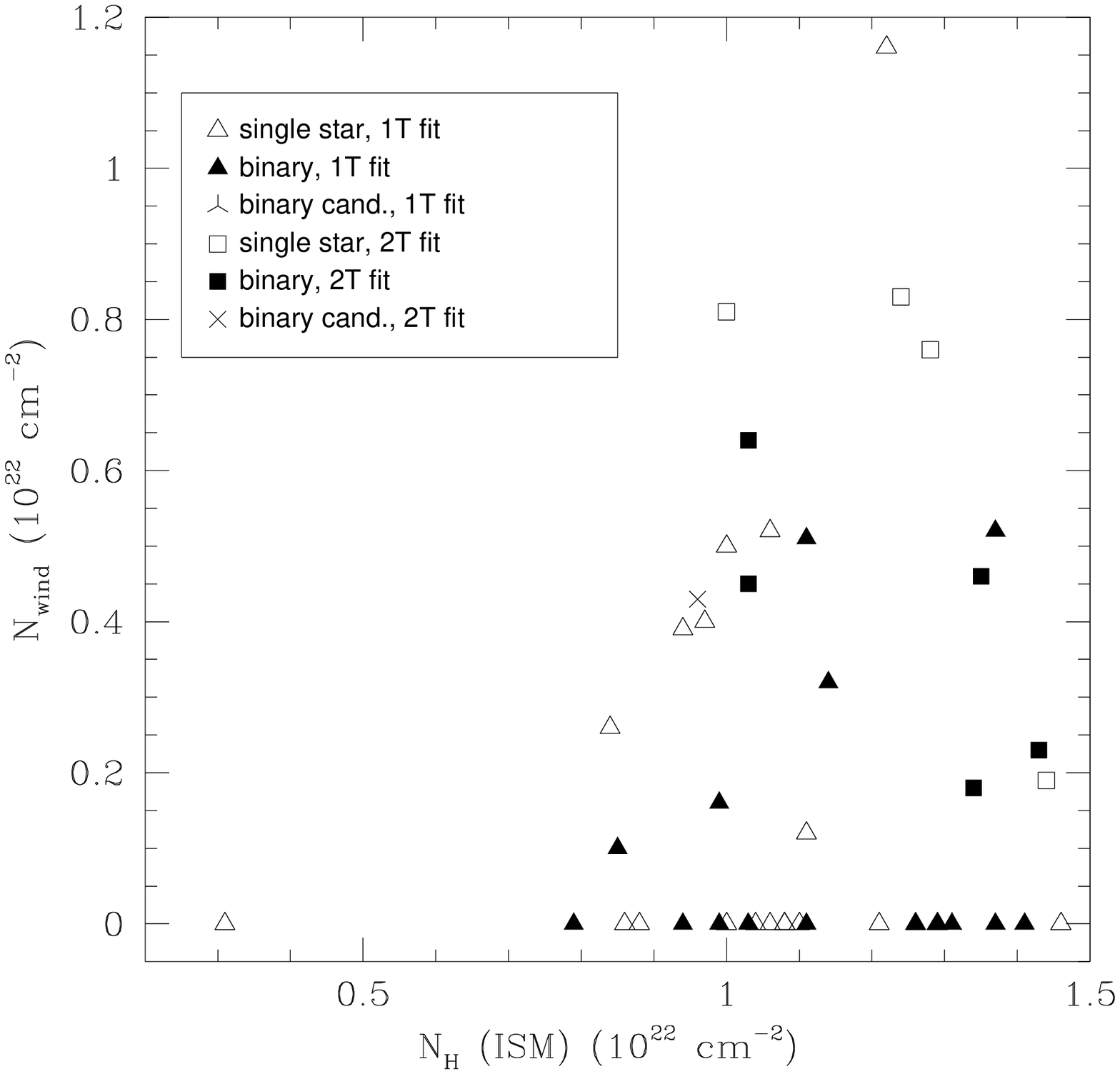}}
\end{center}
\end{minipage}
\hfill
\begin{minipage}{8cm}
\begin{center}
\resizebox{8cm}{!}{\includegraphics{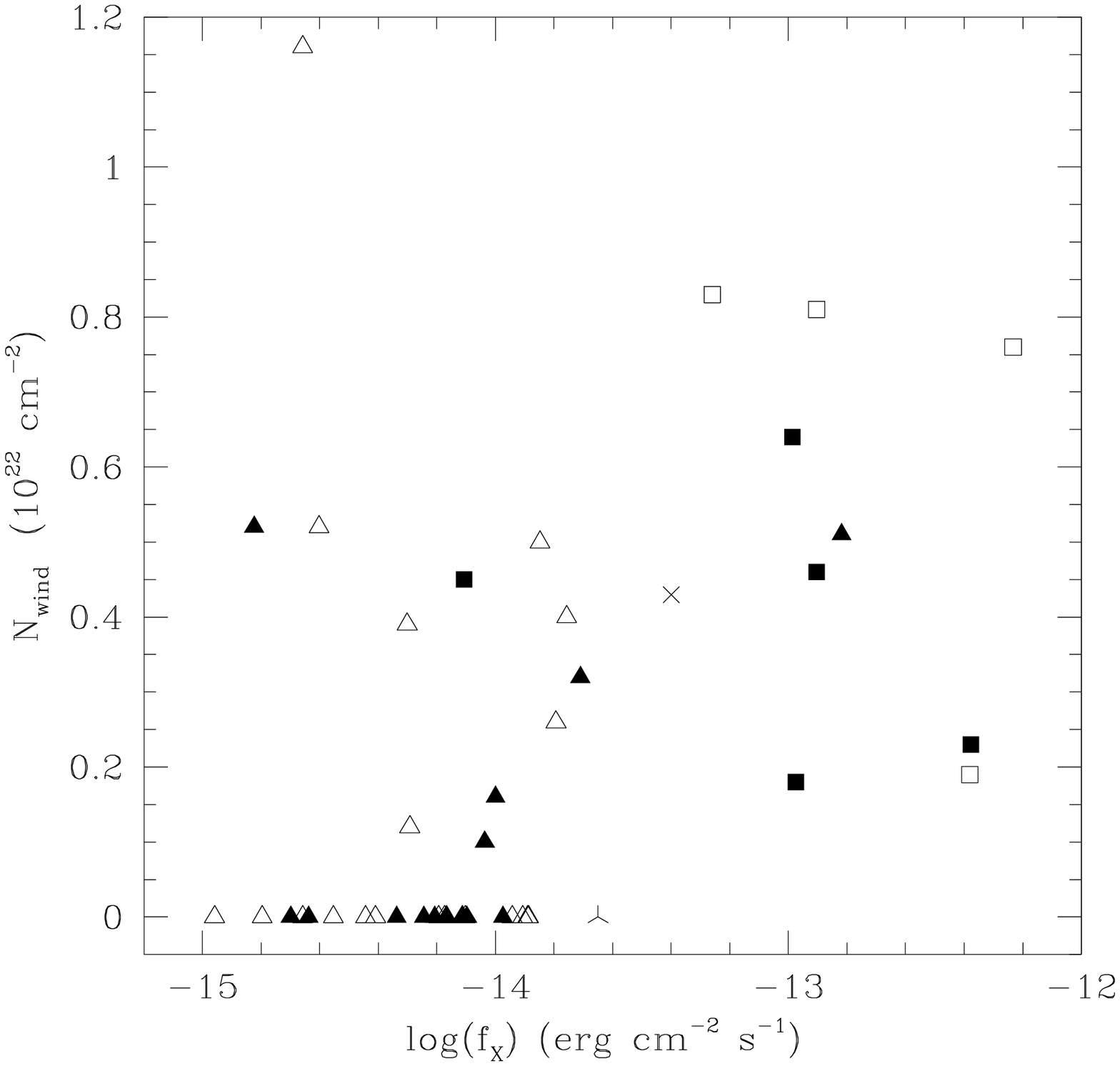}}
\end{center}
\end{minipage}
\caption{Left: best-fit wind column density as a function of interstellar neutral hydrogen column density. Right: best-fit wind column density as a function of observed flux in the 0.5 -- 10\,keV domain. \label{Fig2}}
\end{figure*}

The next brightest sources of our sample are CPR2002\,A20, CPR2002\,A11, MT91\,516, Cyg\,OB2 \#3, and WR\,146. The former and the latter two stars fall outside the field of view covered by the {\it XMM-Newton} observations \citep{Rauw} and we can only estimate the pile-up fraction from the {\it Chandra}-ACIS data themselves. For the two remaining stars, we proceeded in the same way as for the four brightest objects, relying on the parameters derived by \citet{Rauw} for the conversion from {\it XMM-Newton} data. In this way, we estimate pile-up fractions of 3 -- 10\% for these five stars, except for ObsID\,10961 where the actual data indicate a 15\% pile-up fraction for MT91\,516. We keep these objects in our analysis, although their spectra are probably affected by a moderate level of pile-up. As a test, we have repeated the spectral fits, using the {\tt pileup} model implemented in {\tt xspec} and based on the work of \citet{Davis}. The best-fit parameters are usually in good agreement with those obtained without pile-up corrections. The fluxes (both observed and absorption corrected) are on average 5 -- 6\% larger than without pile-up correction. Because of some degeneracy between the plasma parameters and the pile-up model, some of the individual spectra yield larger deviations. We will thus include both solutions (with pile-up correction and without) when discussing the variability of these five sources. 

All other O-stars in our sample should have pile-up fractions of less than 5\%, and we did not apply a correction in these cases. 

\section{Spectral analyses \label{global}}
\subsection{O-type stars}
Excluding the three O-stars with heavy pile-up (Cyg\,OB2 \#8a, \#5 and \#9), we are left with a sample of 49 O-type stars. The fits of the spectra of 10 objects required two plasma components with different temperatures, whilst the remaining spectra were well fitted with a single plasma component. The distributions of the best-fit temperatures and wind column densities are shown in Fig.\,\ref{stats}. 

For single temperature model fits, there is a continuous distribution of $kT$ from about 0.1 to 1.4\,keV, with a prominent peak between 0.5 and 0.6\,keV. This looks quite different from the temperature distribution found for the O-stars in the Carina Nebula \citep{Carina}, which is rather flat between 0.1 and 0.7\,keV and lacks objects with higher values of $kT$. However, for CCD spectra with a low number of counts, there is a well-known degeneracy between the column density and the plasma temperature. In the present case, the fact that we have found a significant number of objects that apparently lack a wind column density in the best-fit model (see below) could bias $kT$ towards higher temperatures. To test this hypothesis, we also show in Fig.\,\ref{stats} the distribution of $kT$ for those objects where the best-fit wind column density is different from zero. This distribution actually lacks the plasma temperatures between 0.5 and 0.6\,keV that dominate the histogram of the entire data set. This confirms that most of the higher best-fit plasma temperatures are associated with a zero wind column density. For 2-T fits, the lower temperature is generally below 0.3\,keV with a few exceptions. The second temperature spans a wide range of values, between 0.6 and 3\,keV.

For 27 stars out of the 39 fitted with a single plasma component, the best-fit column of the wind absorption is found to be zero. One may thus wonder whether there are biases in our sample against the detection of a wind column. The most obvious candidates for such biases are the strong interstellar absorption towards most objects in Cyg\,OB2 and the low flux level of some of the sources. We have tested both hypotheses. The interstellar column has no clear impact on the detection of additional wind absorption. There are however some hints for a potential bias against such a detection for the lowest flux objects (see Fig.\,\ref{Fig2}).
  
\begin{figure}[hbt]
\begin{center}
\resizebox{8cm}{!}{\includegraphics{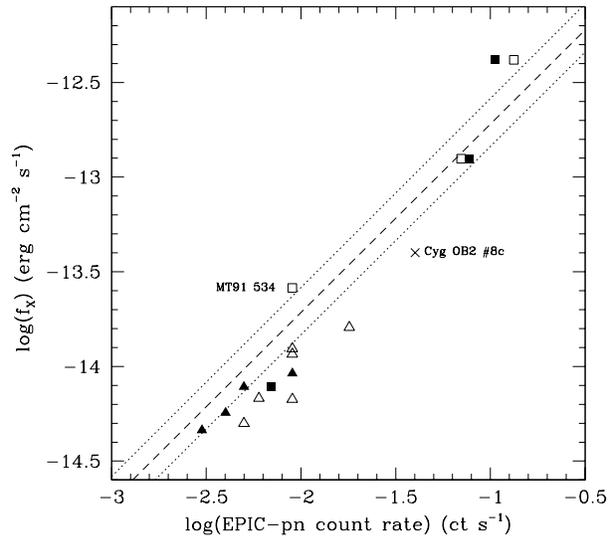}}
\end{center}
\caption{Comparison of the observed fluxes inferred from the ACIS spectra and the count rates of the EPIC-pn instrument aboard {\it XMM-Newton} \citep[from][]{Rauw}. The symbols have the same meaning as in Fig.\,\ref{Fig2}. The dashed straight line yields the conversion factor for a spectral model with the parameters that are the mean values of those inferred here (N$_{\rm H}$(ISM) = $1.07 \times 10^{22}$\,cm$^{-2}$, N$_{\rm H}$(wind) = $0.12 \times 10^{22}$\,cm$^{-2}$, kT = 0.56\,keV). The dotted lines correspond to conversion factors for models with one standard deviation about the mean spectral parameters.\label{ACISpn}}
\end{figure}

Fig.\,\ref{ACISpn} shows a comparison of the observed fluxes inferred from the ACIS spectra analyzed here with the {\it XMM-Newton} EPIC-pn count rates reported by \citet{Rauw} for 17 objects in common. 
Generally speaking, we find that the EPIC-pn count rates of most objects are larger than expected from the fluxes that we infer from the ACIS spectra. In some cases, the difference amounts to a factor two (0.3\,dex). We have also compared the fluxes obtained here with those derived by \citet{YN} using the 2XMM catalogue. The fluxes of the brighter sources are in good agreement, whereas there are rather large discrepancies for the fainter objects.
Whilst some of the discrepancies could be due to variability (see below), it seems extremely unlikely that all stars would be fainter at the time of the {\it Chandra} observations. 

The current status of the cross-calibration of the {\it Chandra}-ACIS and {\it XMM-Newton}-EPIC instruments is discussed by \citet{Schellenberger}. These authors compare the fluxes inferred from a sample of more than 50 clusters of galaxies that are fitted with optically thin thermal plasma models, as are our stellar spectra. These authors find that ACIS yields lower flux than EPIC, mainly in the lower energy range (0.7 -- 2.0\,keV) with typical differences of about 10\%. This is less than the differences found here, but one has to bear in mind that \citet{Schellenberger} focus on {\em extended sources in an otherwise rather empty environment}, whereas our study is concerned with {\em point sources in a crowded environment}. Therefore, the most likely reasons for the larger discrepancies in our case are source confusion in crowded regions with {\it XMM-Newton}, due to its coarser point spread function (PSF), and photon loss due to pile-up of the brighter sources for {\it Chandra}. We have inspected the impact of applying a wider extraction region (such as required for {\it XMM-Newton} data) to the treatment of the ACIS data. Compared to a 2.5 arcsec extraction radius (as used typically for the ACIS data), an extraction radius of 28 arcsec leads to a 50 -- 100\% increase in the number of photons associated with a moderately bright source. This is due to contamination by a weak diffuse emission and/or a conglomerate of faint point sources. The situation is even worse for sources such as Cyg\,OB2 \#8c where the extraction region in the {\it XMM-Newton} data is affected by the wings of the PSF of the very bright Cyg\,OB2 \#8a. In summary, we conclude that most of the discrepancies seen on the fainter sources indeed stem from a contamination of the EPIC spectra by neighboring sources.

\subsection{The L$_{\rm X}$/L$_{\rm bol}$ relation of O-stars in Cyg\,OB2}
\begin{figure*}
\begin{minipage}{8cm}
\begin{center}
\resizebox{8cm}{!}{\includegraphics{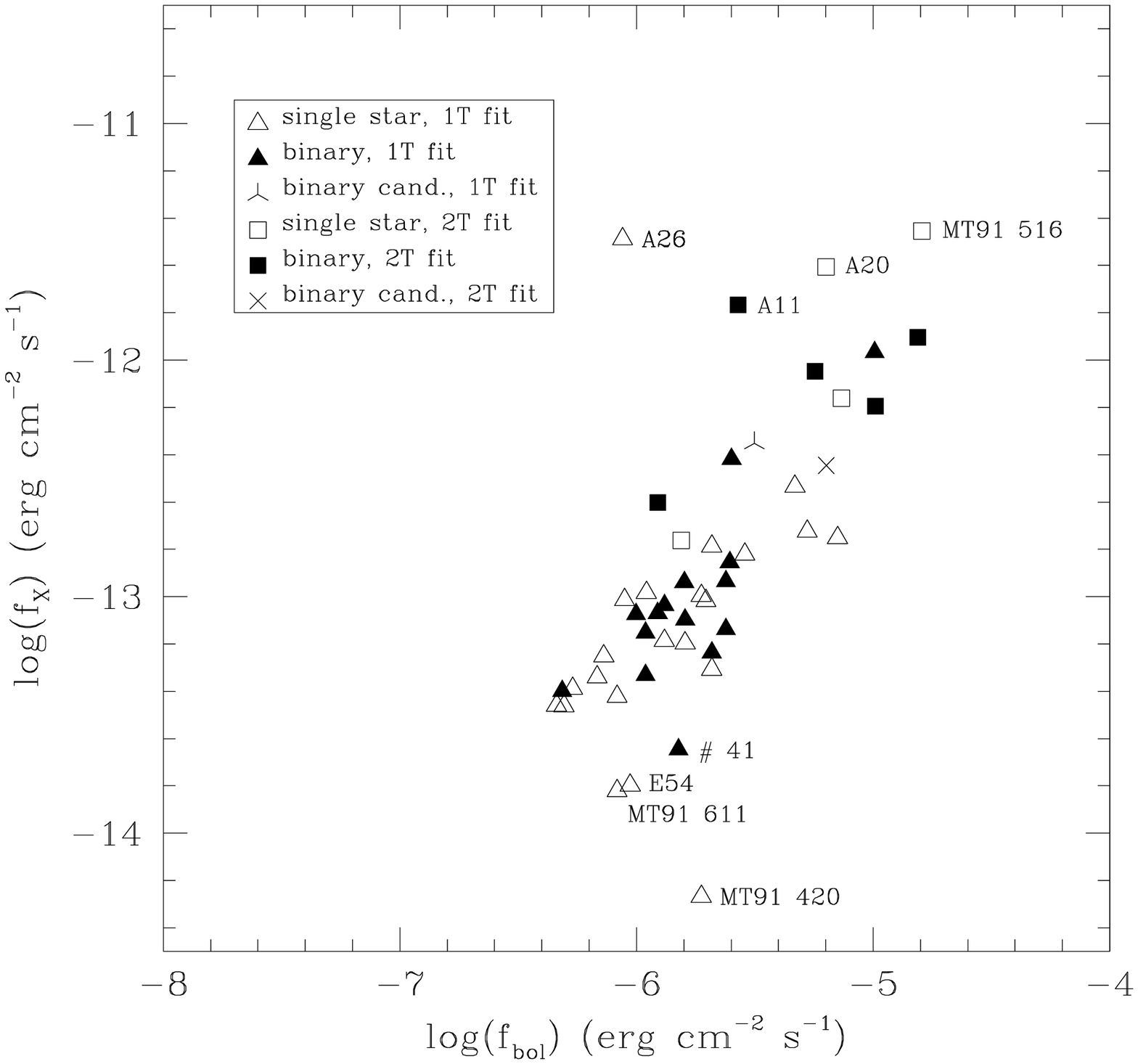}}
\end{center}
\end{minipage}
\hfill
\begin{minipage}{8cm}
\begin{center}
\resizebox{8cm}{!}{\includegraphics{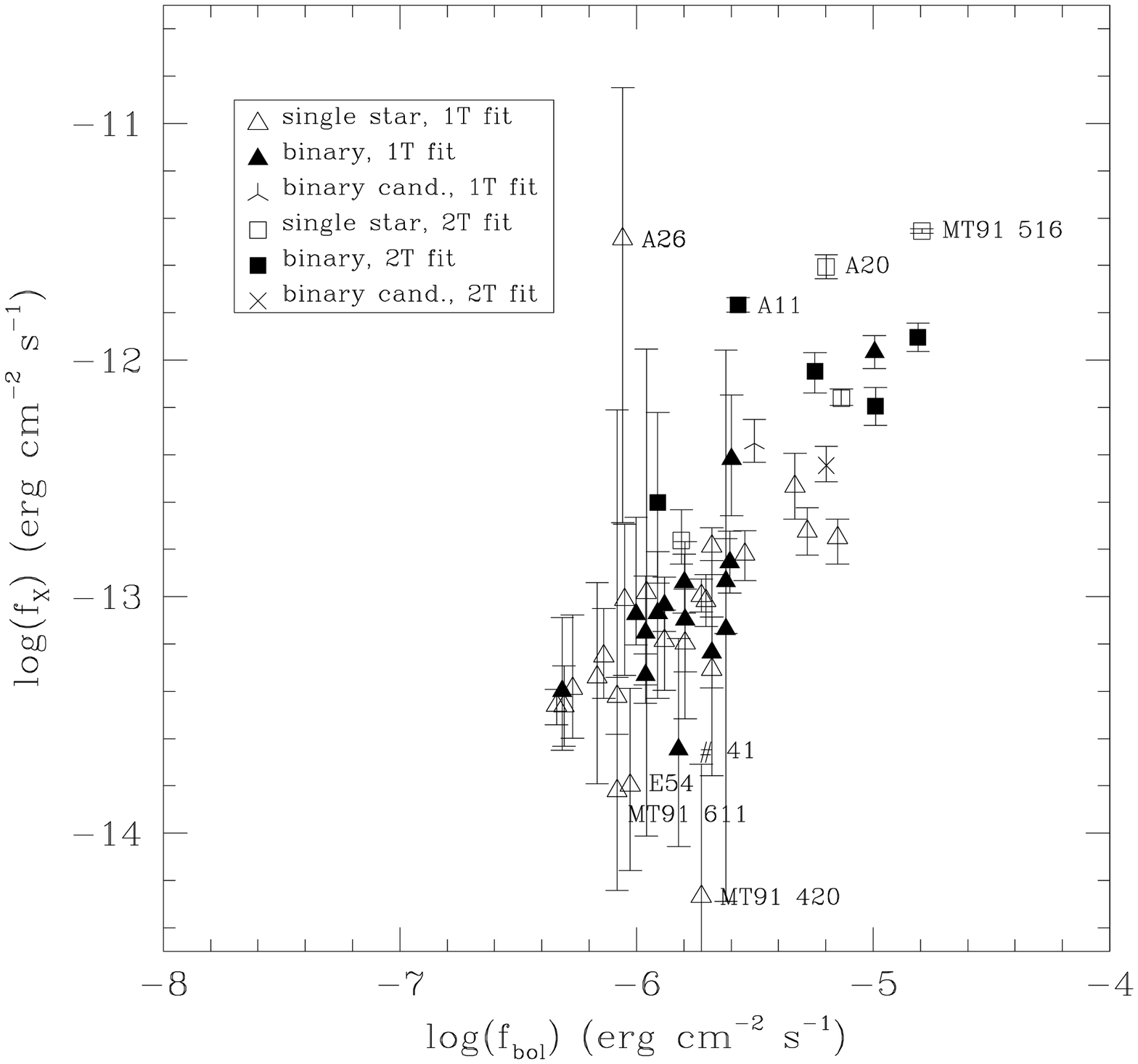}}
\end{center}
\end{minipage}
\begin{minipage}{8cm}
\begin{center}
\resizebox{8cm}{!}{\includegraphics{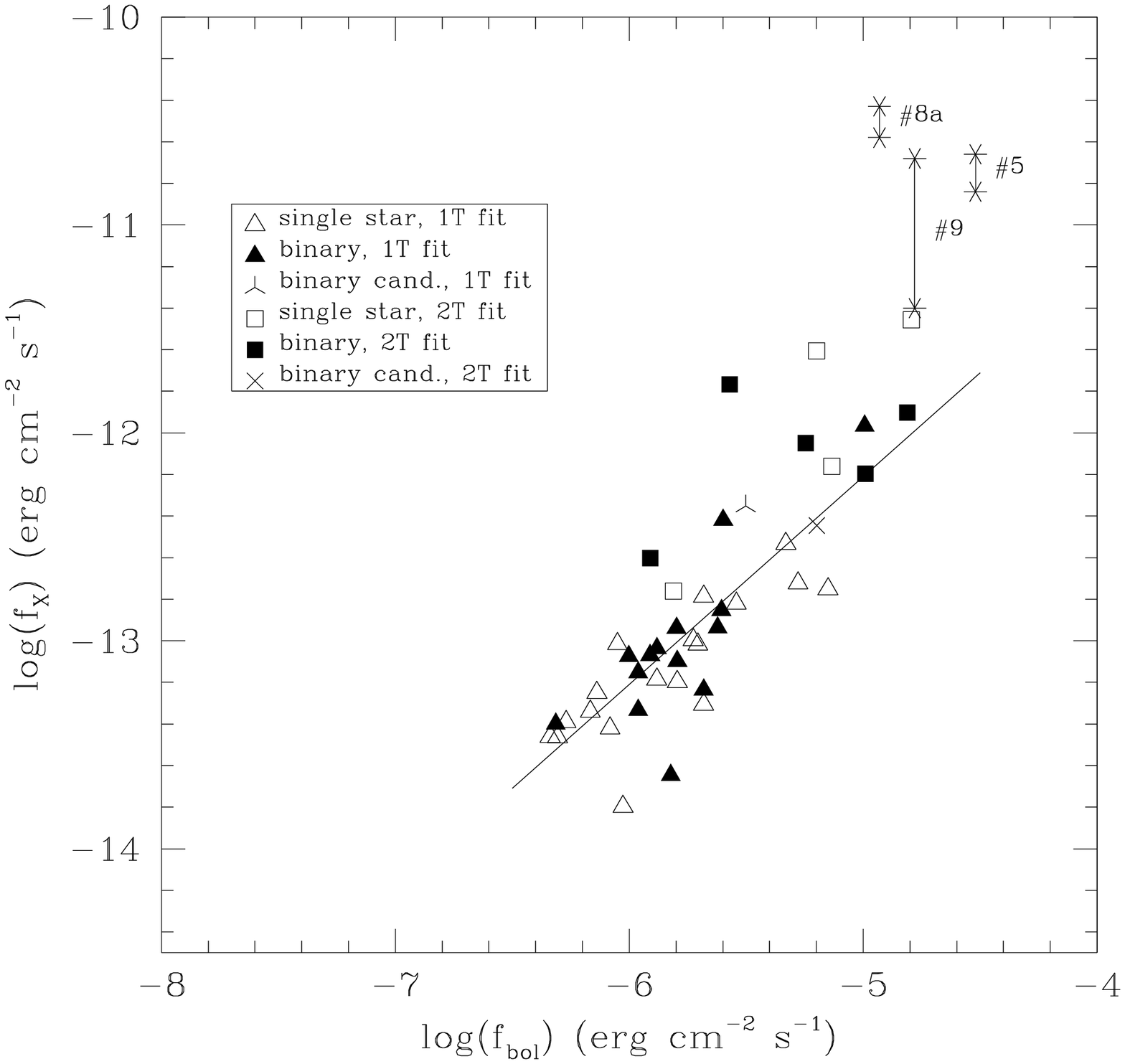}}
\end{center}
\end{minipage}
\hfill
\caption{Top left: relation between the X-ray luminosity and the bolometric luminosity of the O-stars in Cyg\,OB2 as inferred from the ACIS spectra. The X-ray fluxes are evaluated in the 0.5 -- 10\,keV domain and are corrected for the interstellar absorption. Top right: same, but with error bars derived via the {\tt cflux} tool of {\tt xspec}. Bottom left: same as top left, but this time restricting ourselves to results for spectra with more than 30 counts. The straight line corresponds to the scaling relation given by Equation (\ref{eq1}). For comparison, the ranges of X-ray fluxes of the multiple systems Cyg\,OB2 \#5, 8a and 9 as measured with {\it XMM-Newton} are also shown by the asterisks.\label{fxfbol}}
\end{figure*}

To investigate the relation between X-ray luminosity and bolometric luminosity, we rely on the X-ray fluxes and bolometric fluxes, which have the advantage of being independent of the distance of Cyg\,OB2. The X-ray fluxes were inferred from the best-fit models of the ACIS spectra. These fluxes were corrected for the interstellar absorption only, i.e.\ they were not corrected for additional wind absorption. Errors on the absorption corrected fluxes were estimated via the {\tt cflux} command in {\tt xspec}. The bolometric fluxes were computed from the $V$ magnitudes, assuming a value of $R_V = 3.1$ and adopting the bolometric corrections of \citet{MP}. When no $B$, $V$ photometry was available, we relied on the near-IR photometry of \citet{Negueruela}. The resulting relation between X-ray flux and bolometric flux is shown in Fig.\,\ref{fxfbol}. This plot reveals the well-known empirical scaling relation between the bolometric and X-ray luminosities of O-type stars \citep[see][and references therein]{YN}.

Early-type binaries are a priori expected to be more luminous in X-rays as a result of the violent wind-wind interactions \citep[e.g.][]{SBP}. However, Fig.\,\ref{fxfbol} does not show a clear overluminosity of the known binary systems in Cyg\,OB2\footnote{We note however that this statement does not apply to the three X-ray brightest O-stars, Cyg\,OB2 \#5, 8a and 9. These three multiple systems are clearly overluminous (see bottom left panel of Fig.\,\ref{fxfbol}), especially at phases/epochs when their X-ray flux reaches its maximum \citep[see also][]{Cazorla}.}. The same conclusion was already reached by \citet{Oskinova1} from a sample of massive binaries observed with {\it ROSAT}, \citet{NGC6231} for O-type stars in NGC\,6231, \citet{YN} for the general sample of O-type stars observed with {\it XMM-Newton}, \citet{Carina} for the O-stars in the Carina complex, and, most recently, by \citet{HM1} for the O-star population of IC\,2944-8 and Havlen-Moffat\,1. In our sample, the strongest over-luminosities are found for CPR2002\,A26, CPR2002\,A20, MT91\,516, and CPR2002\,A11. The corresponding points are labelled in Fig.\,\ref{fxfbol}. The X-ray flux of the first of these objects is almost certainly spurious. Indeed, the ACIS spectrum of CPR2002\,A26 has a very low number of counts and the error bar on the net flux is huge (see Fig.\,\ref{fxfbol}). Its large absorption corrected X-ray flux results from the correction of an apparently low-temperature plasma ($kT = 0.14^{+.14}_{-.05}$\,keV) by the effect of a large interstellar column density ($N_{\rm H} = 1.26 \times 10^{22}$\,cm$^{-2}$). The position of this point should thus be taken with extreme caution. The other outliers have much higher quality spectra and are thus robust. Conversely, there are a number of underluminous stars towards the lower end of the range in bolometric luminosity explored by our sample. These are MT91\,611, 420 and CP2012\,E54 (three presumably single stars with a 1-T fit) and Cyg\,OB2\,\#41 (a binary candidate). The spectra of some of these objects are again of poor quality. We have thus built a new version of the figure, where we have discarded those data points that correspond to spectra with less than 30 counts (lower left panel of Fig.\ \ref{fxfbol}). 

We can then adjust the L$_{\rm X}$/L$_{\rm bol}$ relation of the O-stars in Cyg\,OB2. For this purpose, we exclude the five data points corresponding to spectra with less than 30 counts, as well as the three overluminous systems (CPR2002\,A11, A20 and MT91\,516), and we combine the data of  Cyg\,OB2 \#22A and \#22B (see Sect.\,\ref{sec22}). Let us start with a simple scaling law between the X-ray flux and the bolometric flux (see also Appendix \ref{annexe}). We obtain the following result: 
\begin{equation}\label{eq1}
\log{{\rm L}_{\rm X}/{\rm L}_{\rm bol}} = -7.21 \pm 0.24
\end{equation}
where all 40 data points were given equal weight in the fit. This result is in perfect agreement with the scaling relation found by \citet{Carina} for the O-type stars in the {\it Chandra} Carina Complex Project. If we weight the data according to the square root of the number of counts in the spectrum, we obtain
\begin{equation}\label{eq2}
\log{{\rm L}_{\rm X}/{\rm L}_{\rm bol}} = -7.15 \pm 0.22
\end{equation}
Alternatively, we can also weight the data according to the {\tt cflux} estimated errors on the X-ray fluxes. This time we find  
\begin{equation}\label{eq4}
\log{{\rm L}_{\rm X}/{\rm L}_{\rm bol}} = -7.18 \pm 0.21
\end{equation} 
Both relations are fully consistent with the unweighted result. Keeping the data points with less than 30 counts in the fits does not change the $\log{{\rm L}_{\rm X}/{\rm L}_{\rm bol}}$ relation, but increases its dispersion. Keeping also the overluminous systems further increases the dispersion and leads to a slight increase of $\log{{\rm L}_{\rm X}/{\rm L}_{\rm bol}}$ but still within the uncertainties of Eqs.\,\ref{eq1}, \ref{eq2} and \ref{eq4}. \citet{AC} reported $\log{{\rm L}_{\rm X}/{\rm L}_{\rm bol}} = -7 \pm 1$ for a sample of 26 OB stars in Cyg\,OB2. Their relation is consistent with ours, but with a much larger dispersion than obtained here. 

\citet{YN} obtained $\log{{\rm L}_{\rm X}/{\rm L}_{\rm bol}} = -6.97 \pm 0.20$ for O stars of Cyg\,OB2 in the 2XMM catalogue. This is 0.25\,dex brighter than what we find here. If we restrict the 2XMM sample to those stars that are not known to be overluminous (either from {\it XMM-Newton} data or from the present study), we are left with 20 objects which have $\log{{\rm L}_{\rm X}/{\rm L}_{\rm bol}} = -7.10 \pm 0.26$. The difference is now reduced to 0.1\,dex. This remaining difference can stem from the issues discussed above and/or a slightly different treatment of the bolometric luminosity (e.g.\ differences in adopted spectral types).

\begin{figure}[h]
\begin{center}
\resizebox{8cm}{!}{\includegraphics{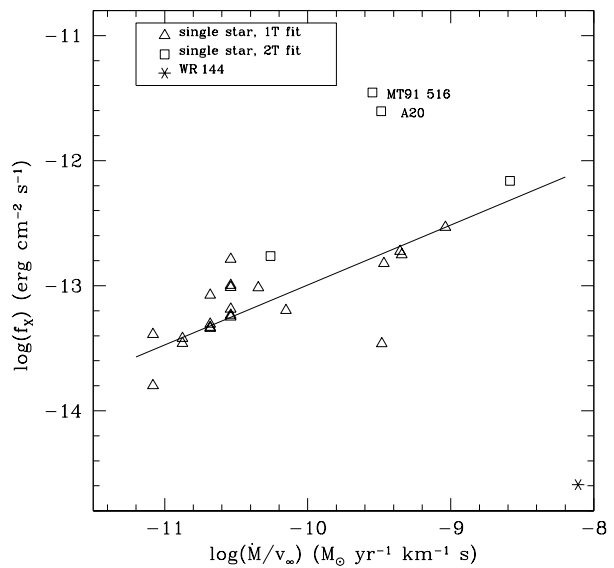}}
\end{center}
\caption{Logarithm of the ISM absorption corrected X-ray fluxes of presumably single O stars, and of WR\,144 as a function of $\log{\frac{\dot{M}}{v_{\infty}}}$.\label{fxMdot}}
\end{figure}

Following the suggestion by \citet{Owocki}, we have searched for a scaling of L$_{\rm X}$ with $\frac{\dot{M}}{v_{\infty}}$. For this purpose, we need to estimate the wind parameters $\dot{M}$ and $v_{\infty}$. We have based our evaluation of these parameters on the mass-loss rates inferred using the \citet{Vink} formalism and tabulated by \citet{Muijres}, and on the assumption that $v_{\infty} = 2.6\,v_{\rm esc}$ \citep{Lamers} for all O-type stars. If we restrict ourselves to the presumably single stars in our sample (weighted according to the estimated errors on the fluxes), and discarding the two overluminous stars MT91\,516 and CPR2002\,A20, we obtain 
\begin{equation}\label{eq3}
\log{{\rm f}_{\rm X}} = (0.48 \pm 0.10)\,\log{\frac{\dot{M}}{v_{\infty}}} - 8.19 \pm 0.97
\end{equation}
This result is shown in Fig.\,\ref{fxMdot}. In terms of the \citet{Owocki} relation, the X-ray luminosity should scale with $(\frac{\dot{M}}{v_{\infty}})^{1-m}$ over most of the spectral range of O-type stars, where the LDI shocks that produce the X-rays are radiative and the winds themselves are optically thin. We thus find that $m \simeq 0.52 \pm 0.10$, which is slightly larger than the upper limit of the range (0.2 -- 0.4) proposed by \citet{Owocki}. We note however that there is considerable scatter around the relation given by equation\,\ref{eq3}. This could indicate that either the relation is not tight, or that the values of the wind parameters are not well enough determined with the assumptions made here.

Our current results do not show a change in the $\log{{\rm L}_{\rm X}/{\rm L}_{\rm bol}}$ relation towards the higher or lower luminosity end of the O-star range, as was suggested by \citet{Owocki}. At the lower luminosity end, we actually find that the $\log{{\rm L}_{\rm X}/{\rm L}_{\rm bol}}$ relation of O-stars holds also for the most luminous B-stars (see Sect.\,\ref{Bstars}). Concerning the high luminosity end, it must be stressed though that the Wolf-Rayet star WR\,144 clearly deviates from the relation in Fig.\,\ref{fxMdot}. Furthermore, the O-star population of Cyg\,OB2 does not contain O\,If$^+$ supergiants which are likely transition objects between O and WR stars, and would thus be the ideal targets to search for the predicted change in the $\log{{\rm L}_{\rm X}/{\rm L}_{\rm bol}}$ relation at the highest luminosities \citep{DeB}.

\begin{figure}[h]
\begin{center}
\resizebox{8cm}{!}{\includegraphics{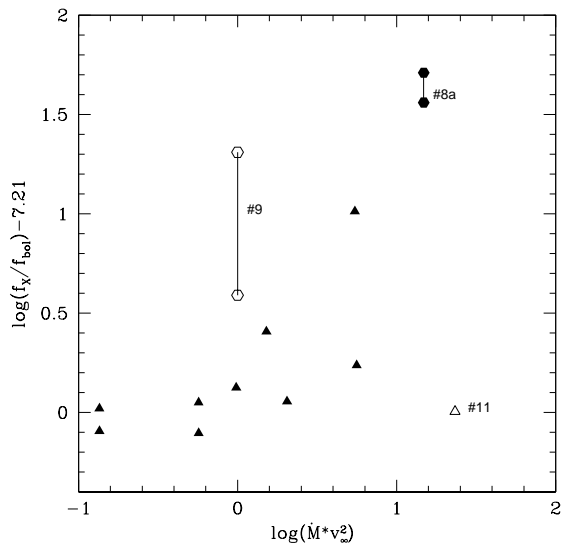}}
\end{center}
\caption{Overluminosity in the X-ray domain versus kinetic power of the primary wind for the known binary systems in our sample. Triangles indicate {\it Chandra} ACIS-I data, whilst the data for Cyg\,OB2 \#8a and 9 (hexagons) are taken from {\it XMM-Newton} observations. Filled symbols stand for systems with orbital periods of less than 30\,days.\label{CWB}}
\end{figure}

As pointed out above, we find no indication of a strong X-ray overluminosity of the known binary systems in the ACIS data. Yet it is interesting to consider this situation in a more detailed way. Wind interactions in relatively short period O + OB systems are usually in the radiative regime where the X-ray luminosity of the wind-wind collision scales with $\dot{M}\,v_{\infty}^2$. If we consider the known O + OB binary systems (leaving the higher multiplicity system Cyg\,OB2 \#5 aside) with orbital periods shorter than 30\,days, we find indeed a roughly linear increase of the X-ray overluminosity with the kinetic power of the primary star wind, although the scatter is quite substantial (see Fig.\,\ref{CWB}). The two long-period systems Cyg\,OB2 \#9 ($P = 860$\,days) and \#11 ($P = 72$\,days), do not follow this trend. \citet{Naze9} showed that the wind collision zone in Cyg\,OB2 \#9 is indeed in the adiabatic regime, where the X-ray luminosity scales with $\dot{M}/(v_{\infty}\,d)$, with $d$ being the orbital separation. Given its orbital period, a similar situation probably applies to Cyg\,OB2 \#11. 

\begin{figure}
\begin{center}
\resizebox{7cm}{!}{\includegraphics{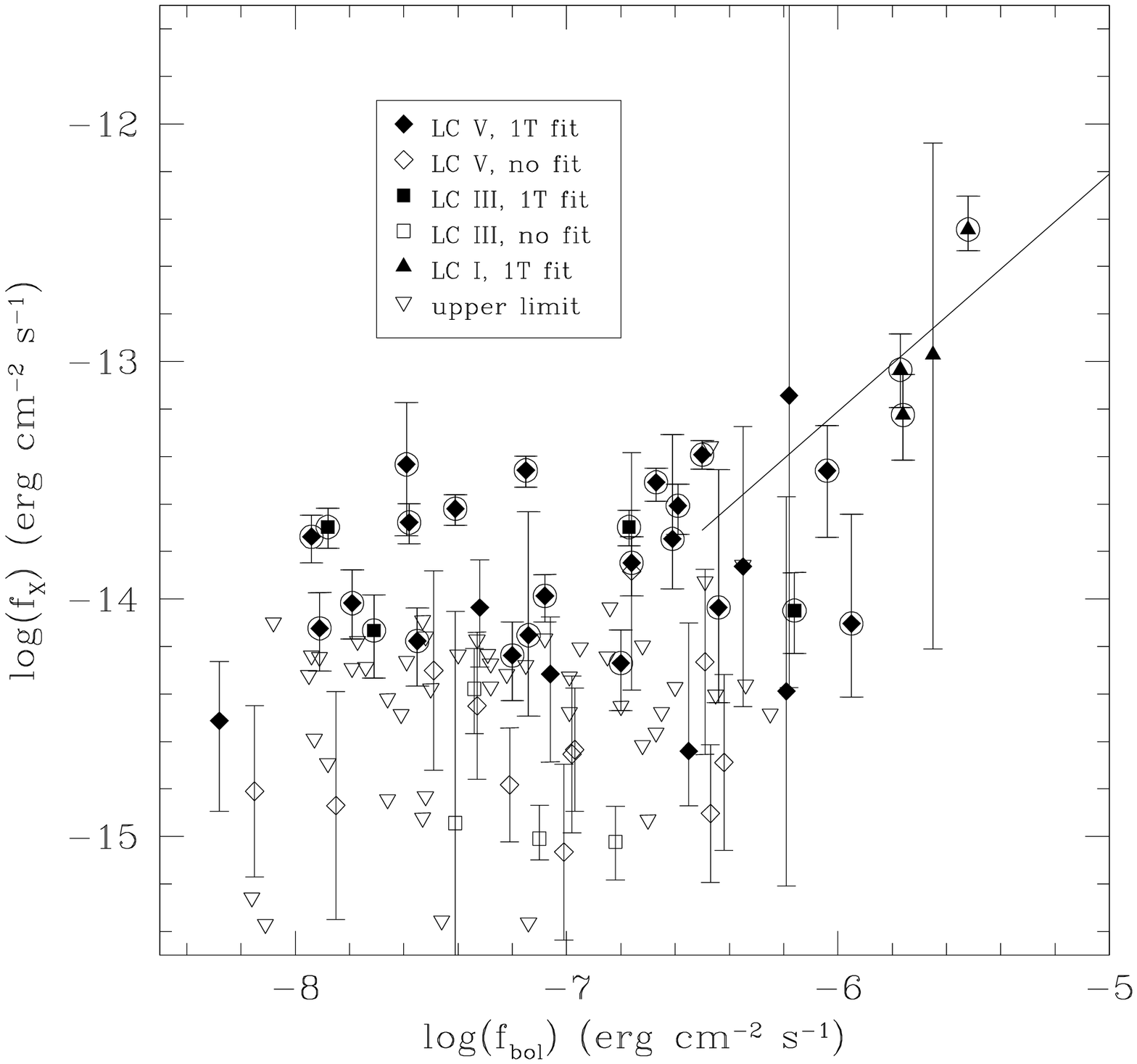}}
\end{center}
\begin{center}
\resizebox{7cm}{!}{\includegraphics{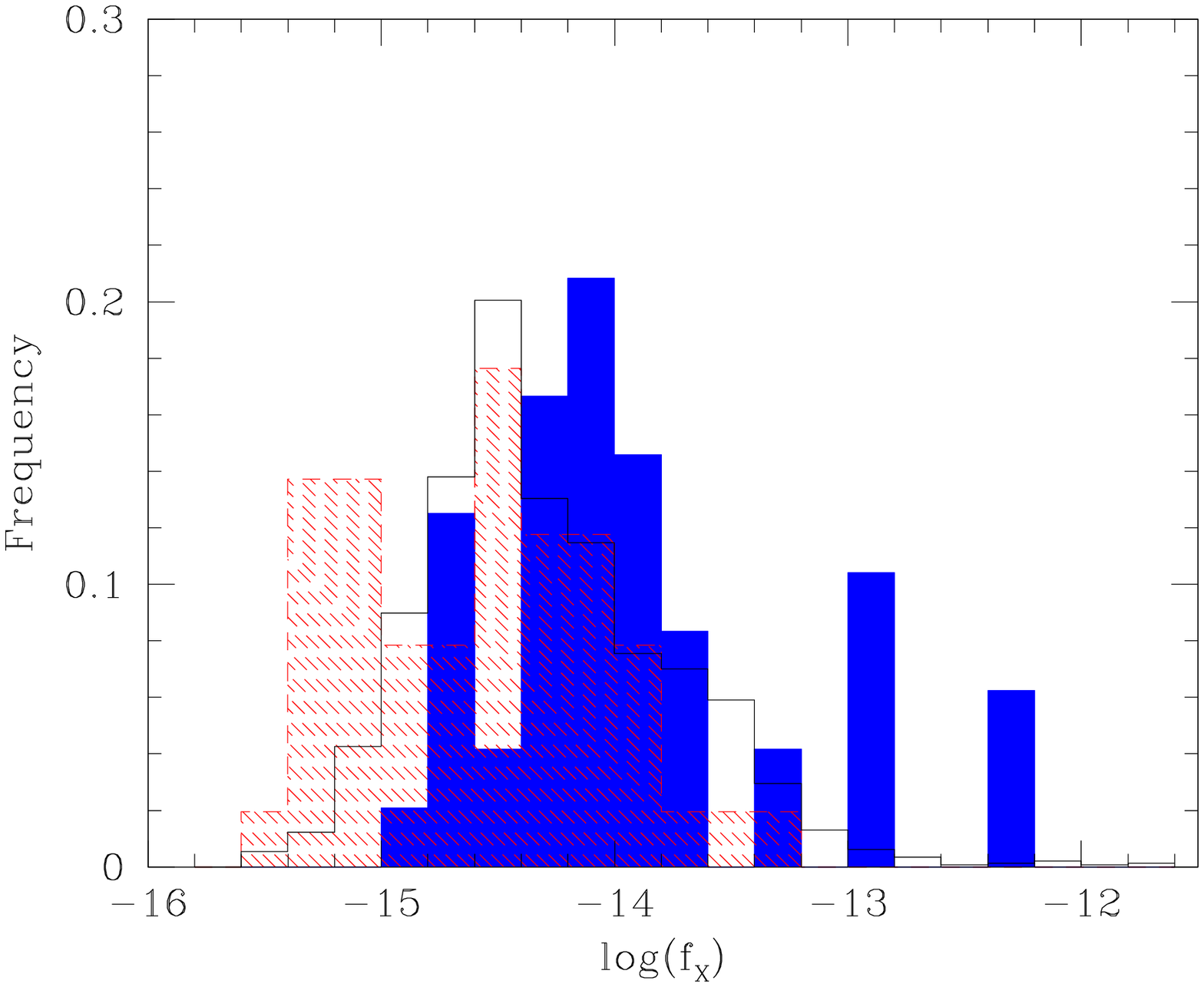}}
\end{center}
\caption{Top: ISM corrected X-ray fluxes of the B-type stars as a function of their bolometric fluxes. The different symbols identify the luminosity classes of the stars and indicate whether or not the fluxes were derived from a spectral fit. Symbols corresponding to objects with more than 30 counts in their spectra are encircled. The straight line indicates the best-fit relation for O-type stars. Bottom: distribution of the observed fluxes of B-type stars (hatched red histogram), O-type stars (blue filled histogram) and the full population of 1457 Cyg\,OB2 X-ray sources from \citet{WD} cleaned for foreground and background sources (black empty histogram).
\label{fxfbolB}}
\end{figure}

\subsection{B-stars \label{Bstars}}
Aside from the bright blue hypergiant Cyg\,OB2 \#12 \citep{Rauw,LBV,Cazorla} which suffers from pile-up in the ACIS data, the vast majority of the B-type stars in our sample are rather faint X-ray sources. First, we have performed a spectral fit for those sources with a sufficient number of counts\footnote{We require again a minimum of four bins in the binned spectra.} in their combined spectra (35 objects out of 51), using the same model as for the O-stars. There are two families in the best-fit parameters. The majority (2/3) of the objects have a spectrum that does not require an absorption component in addition to the ISM column density. The $kT$ of these objects lies between 0.2 and 4.4\,keV, with a mean of $(2.4 \pm 1.3)$\,keV. The remaining objects apparently require additional columns in the range 0.1 to $2.2 \times 10^{22}$\,cm$^{-2}$ with a mean value of $(0.79 \pm 0.65)\times 10^{22}$\,cm$^{-2}$. The corresponding temperatures are lower (between 0.1 and 1.6\,keV), suggesting that the additional columns are due to the above-mentioned degeneracy between temperature and column density in the fits of X-ray spectra with a low number of counts. The X-ray plasma in the B-type stars appears to be generally hotter than in the O-type stars, although there is a large dispersion in $kT$ for the B-stars. The lowest temperatures are usually found to be associated with early-type B-stars, although some late-type B-stars also have low $kT$ and some early-type B-stars are found to have large $kT$. 

We have used the above fits to derive X-ray fluxes corrected for the ISM absorption. For the 16 objects where no spectral fit could be achieved, we have derived observed and ISM absorption-corrected fluxes, assuming that their count rates are consistent with a spectrum described by a thermal plasma model with $kT = 2.4$\,keV, absorbed by the sole ISM column. The resulting distribution of X-ray fluxes versus bolometric fluxes is shown in Fig.\,\ref{fxfbolB}. The bolometric fluxes were taken from \citet{Wright}. For those known B-stars that were not detected in X-rays, we have evaluated upper limits on the X-ray luminosities. For this purpose, 1-$\sigma$ upper limits on the number of detected counts were estimated by inserting their position into the ACIS-extract pipeline. These were then converted into photon fluxes following relation (8) of \citet{Broos}, and into ISM absorption-corrected fluxes assuming $kT = 2.4$\,keV and accounting for the ISM column density. The results are shown by the downwards pointing open triangles in Fig.\,\ref{fxfbolB}.

The relation valid for O-stars that we have derived above, still holds for the brightest B-stars (early B supergiants with $\log{\frac{L_{\rm bol}}{L_{\odot}}} \geq 4.9$, corresponding here to $\log{f_{\rm bol}} \geq -5.9$). For fainter stars, there is a huge dispersion. Restricting ourselves to the brightest objects (encircled in Fig.\,\ref{fxfbolB}) would suggest a flattening of the  L$_{\rm X}$/L$_{\rm bol}$ relation \citep[see the case of NGC\,6231,][]{NGC6231}. This feature as well as the generally high plasma temperatures of B-stars could reflect X-rays from magnetically confined winds \citep{BabelMontmerle}. However, studying a sample of early-type B-stars with known magnetic fields, \citet{OskinovaB} found that hard and strong X-ray emission does not necessarily correlate with the presence of a magnetic field. Moreover, such a flattening is not supported by the full sample. The same conclusion was reached from the data of the {\it Chandra} Carina Complex Project \citep{Carina}. 


The X-ray flux of the brightest non-supergiant B stars reaches about $4 \times 10^{-14}$\,erg\,cm$^{-2}$\,s$^{-1}$ which, adopting a distance of 1.4\,kpc, corresponds to an X-ray luminosity of about $10^{31}$\,erg\,s$^{-1}$. The latter is not inconsistent with the possibility that the X-ray emission arises from a low-mass pre-main sequence companion. In this context, the bottom panel of Fig.\,\ref{fxfbolB} compares the distribution of observed fluxes of O and B stars with the distribution for the full set of Cyg\,OB2 sources from \citet{WD} cleaned for foreground and background sources \citep[see][]{Wright2010}. The latter should be dominated by low-mass pre-main sequence stars. \citet{Gagne} performed a similar exercise for the B-stars in the Chandra Carina Complex Project and found an excess of X-ray bright B-stars compared to the distribution of PMS stars. Fig.\,\ref{fxfbolB} does not show such an excess and hence does not argue against the low-mass companion scenario. 

\subsection{Wolf-Rayet stars}
\paragraph{WR\,144}
WR\,144 is a presumably single WC4 star \citep[][and references therein]{Sander}. With a net number of 5.8 counts, this star corresponds to the weakest detection of a WR star in our sample. To the best of our knowledge, this is the first X-ray detection of a presumably single WC star. Our spectrum has only two energy bins, which is not sufficient to perform a decent spectral fit. The energies of the source counts indicate a hard emission, probably as a result of a heavy circumstellar absorption. Although the data are not of sufficient quality for a spectral fit, we can use them to obtain some estimate of the X-ray flux. For this purpose we have assumed a single-temperature model with the plasma composition as derived by \citet{Sander} absorbed by the ISM column and an additional wind column with the same composition as the emitting plasma. We built a grid of models with $kT$ varying between 0.3 and 3.0\,keV where the only variables correspond to the normalization parameter of the {\tt vapec} component, and the wind column density. To within a factor 1.5 uncertainty, we find that the detected photons correspond to a flux (corrected for ISM absorption) of $\sim 2.55 \times 10^{-15}$\,erg\,cm$^{-2}$\,s$^{-1}$. Comparing with the bolometric flux derived from the bolometric luminosity inferred by \citet{Sander} yields $\log{L_{\rm X}/L_{\rm bol}} = -8.8 \pm 0.2$. This low value is consistent with previous results on single WC stars: \citet{Oskinova} reported the non-detection of the WC5 star WR\,114 and argued that single WC stars are X-ray faint with $\log{L_{\rm X}/L_{\rm bol}} \leq -8.4$ for WR\,114. Our result qualitatively and quantitatively fits into this picture.

\paragraph{WR\,145}
With a spectral type WN7/CE + O7\,V((f)), WR\,145 ($\equiv$ MR\,111) is one of a few WR stars in our Galaxy with a hybrid WN/WC spectral type. \citet{Sander} found that its spectrum is well fitted by a WN-type model with enhanced carbon. WR\,145 is a spectroscopic binary with a period of 22.5\,days showing evidence for a wind-wind interaction in the profiles of optical WR emission lines \citep[see][and references therein]{Muntean}. 

\begin{figure*}
\begin{minipage}{8cm}
\begin{center}
\resizebox{8cm}{!}{\includegraphics{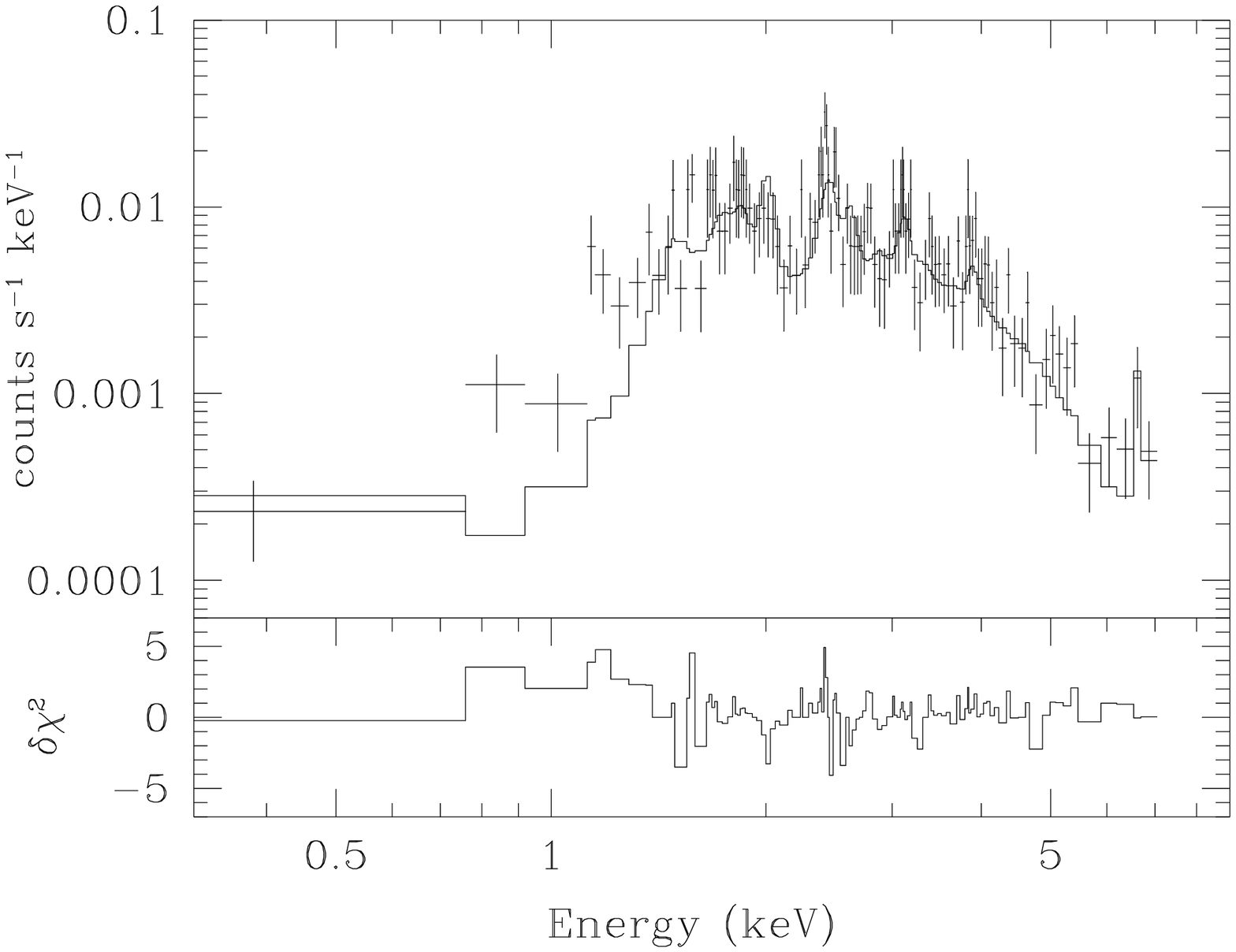}}
\end{center}
\end{minipage}
\hfill
\begin{minipage}{8cm}
\begin{center}
\resizebox{8cm}{!}{\includegraphics{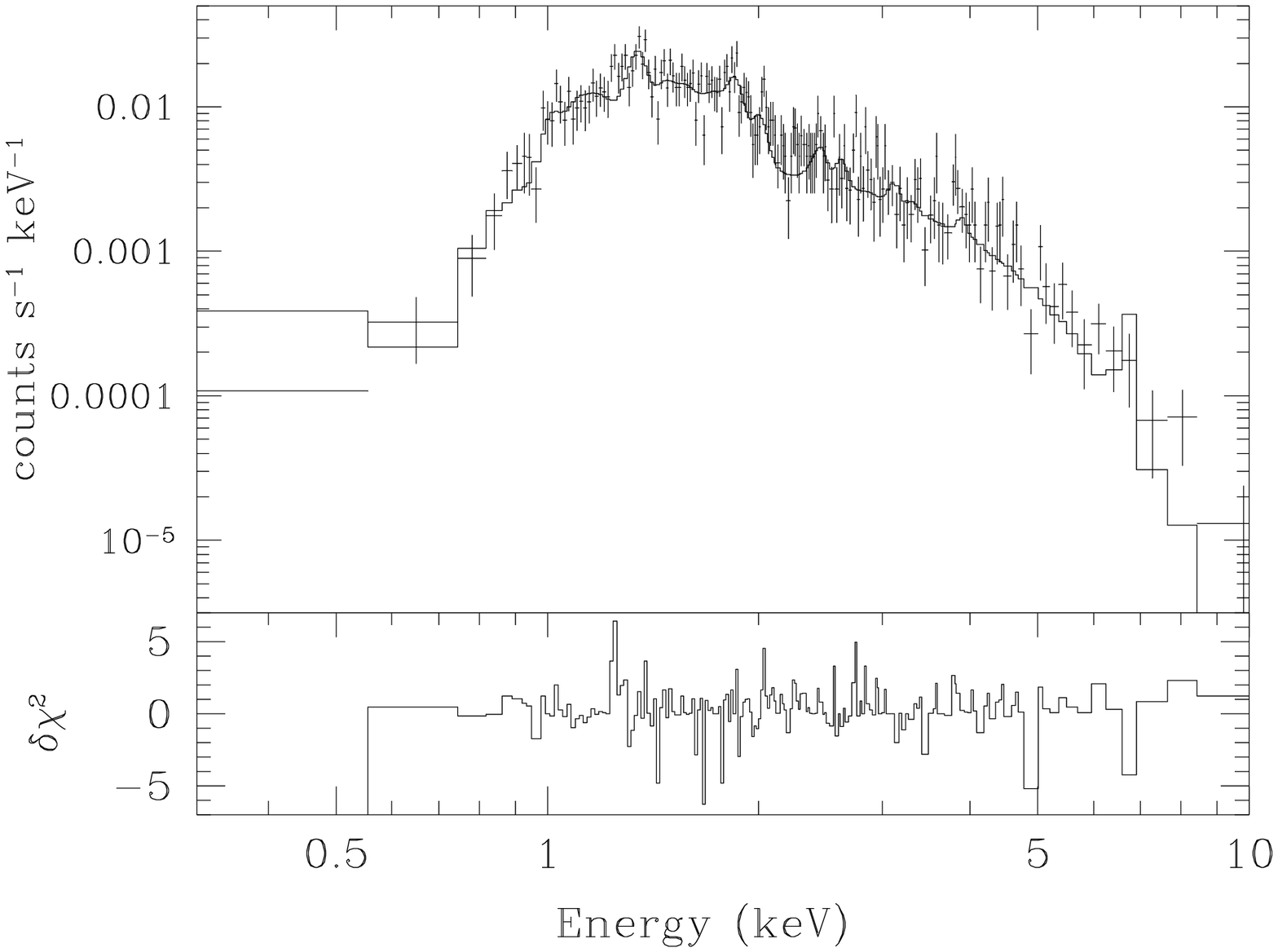}}
\end{center}
\end{minipage}
\caption{ACIS-I spectra of WR\,145 (left) and WR\,146 (right), along with their best fit assuming a non-solar composition for WR\,145 and solar abundances for WR\,146.\label{WRspec}}
\end{figure*}

As far as its observed X-ray emission is concerned, WR\,145 is the brightest WR star in Cyg\,OB2. Its X-ray emission was already detected with {\it EINSTEIN} \citep{Pollock} and {\it ROSAT} \citep{RASS}. The {\it EINSTEIN}-IPC data yield a count rate between 3 and $9 \times 10^{-3}$\,cts\,s$^{-1}$, although the differences between the different pointings are not statistically significant \citep{Pollock}. {\it ROSAT} observed WR\,145 twice, once during the All-Sky Survey (PSPC-C count rate of $(7.4 \pm 3.5) \times 10^{-3}$\,cts\,s$^{-1}$) and once during a pointed observation \citep[PSPC-B count rate of $(2.8 \pm 2.5) \times 10^{-3}$\,cts\,s$^{-1}$,][]{RASS}. In the {\it Chandra} survey, the star was only observed once at orbital phase $0.34$ according to the ephemerides of \citet{Muntean}. Our data thus correspond to an orbital phase shortly after quadrature with the O-star companion being closer to us than the WR star. We have fitted the spectrum with a model\footnote{\citet{Zhekov} analyzed the same spectrum along with several archival {\it XMM-Newton} spectra and concluded that a 2-T plasma model was necessary to represent the data ($kT_1 = 0.99$, $kT_2 = 4.8$\,keV). Our fits do not require a second plasma component.} of the kind {\tt phabs*vphabs*vapec} (see Fig.\,\ref{WRspec}) where the ISM column density was set to $9.7 \times 10^{21}$\,cm$^{-2}$ and the non-solar abundances were adopted, both for the emitting plasma ({\tt vapec}) and the circumstellar absorption ({\tt vphabs}), from \citet{Sander}. Abundances (with respect to hydrogen) of He, C and O were frozen at 1000 times solar. The N abundance was set to 0.001 times solar and all other elements were set at 706 times solar. A good fit ($\chi_{\nu}^2 = 1.08$ for 110 degrees of freedom) is achieved with a single temperature model with $N_{\rm wind} = (3.4^{+.6}_{-.9}) \times 10^{19}$\,cm$^{-2}$ and $kT = 1.59^{+.38}_{-.17}$\,keV. The comparatively low value of $N_{\rm wind}$ indicates that the bulk of the X-ray emission probably arises from either the wind-wind interaction zone or the O7\,V companion. In this case, and given the orbital phase of our observation, one could argue that the circumstellar column towards the hot plasma might have a roughly solar abundance. In fact, adopting solar abundances for the wind column yields an equal quality fit, but this time with $N_{\rm wind} = (2.82^{+.52}_{-.71}) \times 10^{22}$\,cm$^{-2}$. We have further tested a model with solar composition also for the emitting plasma. The fit is actually marginally better ($\chi_{\nu}^2 = 1.05$ for 110 degrees of freedom), and the parameters are $N_{\rm wind} = (2.68^{+.53}_{-.73}) \times 10^{22}$\,cm$^{-2}$ and $kT = 1.60^{+.43}_{-.19}$\,keV. Whatever the adopted abundances, the observed and ISM corrected X-ray fluxes are equal to $5.3 \times 10^{-13}$ and $6.5 \times 10^{-13}$\,erg\,cm$^{-2}$\,s$^{-1}$ respectively. The dereddened flux is a factor 2.5 larger than what we would expect from the sole O7\,V((f)) companion, based on our $\log{\rm L_{\rm X}/{\rm L}_{\rm bol}}$ relation for O-type stars. Therefore, it seems probable that most of the X-ray emission of WR\,145 arises from the O-star companion along with a contribution from the wind-wind interaction.

Our variability test reveals a clear intra-pointing variability of the  X-ray emission of WR\,145 (Kolmogorov-Smirnov statistics of $6 \times 10^{-7}$) during the single observation. \citet{Zhekov} reported some small differences between the absorption-corrected fluxes of the various {\it XMM-Newton} and {\it Chandra} observations. We have folded our best-fit (solar composition) model through the response matrices of the {\it EINSTEIN}-IPC and {\it ROSAT}-PSPC-C instruments. The predicted count rates are $(6.7 \pm 1.1) \times 10^{-3}$\,cts\,s$^{-1}$ for the IPC, and $(1.9 \pm 1.5) \times 10^{-3}$\,cts\,s$^{-1}$ for the PSPC-C. The agreement with the {\it EINSTEIN} data is reasonable, whilst the {\it ROSAT} All Sky Survey observations yields a much larger count rate than predicted here. However, given the large uncertainties (mainly due to the short integration time of the RASS on this source), this difference is not highly significant. The older X-ray data are thus not helpful to clarify the issue of variability of the X-ray flux of WR\,145.

\paragraph{WR\,146}
WR\,146 is a visual binary system consisting of a WC6 star with an O8\,I-IIf companion \citep{Lepine} at a separation of 0.168\,arcsec \citep{Niemela}. Radio observations reveal a thermal component associated with the WC5 star, along with an elongated non-thermal component \citep{Dougherty}. Using high-resolution {\it HST} images and the radio maps of \citet{Dougherty}, \citet{Niemela} demonstrated that the non-thermal radio emission arises in between the two components, thereby establishing this emission as a result of the wind-wind interaction. However, the system might actually be more complex. Indeed, \citet{SG} found variations in the radio light curve of WR\,146 on different time-scales (decades, 3.38\,years and weeks), and interpreted the 3.38\,yr periodicity as evidence for the presence of a third component orbiting the O8 star. A similar conclusion was reached by \citet{Dougherty2} based on the level of radio emission, and the derived mass-loss rate, of the O8 star. These authors suggested that the latter might actually consist of an O8 star orbited by another WC star.

X-ray emission from WR\,146 was previously reported with {\it EINSTEIN} \citep[IPC count rate $6^{+5}_{-.4} \times 10^{-3}$\,cts\,s$^{-1}$,][]{Pollock} and {\it ROSAT} \citep[PSPC-C count rate $(3.5 \pm 1.9) \times 10^{-3}$\,cts\,s$^{-1}$,][]{RASS}. WR\,146 was observed three times with {\it Chandra}, once in March 2007 (JD\,2454177.31) and twice within one day (JD\,2455258.18 and 2455258.53) in the course of the survey in March 2010. As a first step, we have again adopted non-solar abundances from \citet{Sander}, for both the emitting and absorbing gas. In this case, abundances (with respect to hydrogen) of He, C, N and O were set to 897, 1000, 0.001 and 1000 times solar. All other elements were set at 706 times solar. The spectra require a 2T-plasma model to achieve a good quality fit. As the model parameters of the three observations agree within the error bars, we conclude that there is no strong evidence for a time-dependence of the X-ray emission of WR\,146 (see below). We thus focus on the spectrum obtained from the combination of all available ACIS data.

\begin{table*}[h!tb]
\caption{Summary of the variability study based on the photon fluxes corrected for the average response. The last column yields the ObsID of the {\it Chandra} pointing (if any) at which intra-pointing variability is detected.\label{variab}}
\begin{center}
\begin{tabular}{|l l c c c|}
\hline
Star & Spectral type & \multicolumn{3}{c|}{Variability}\\
     &               & Inter-epoch & Intra-pointing & ObsID\\
\hline
\multicolumn{5}{|c|}{O-stars}\\
\hline
CPR2002\,A15 & O7\,I             & N & Y & 12099\\
Cyg\,OB2 \#3  & O6\,IV + O9\,III  & Y & N & -- \\ 
Cyg\,OB2 \#4  & O7\,III           & Y & N & -- \\ 
Cyg\,OB2 \#15 & O8\,V + B         & Y & N & -- \\
CPR2002\,A11         & O7.5\,II + OB     & Y & N & -- \\ 
Cyg\,OB2 \#22 & O3\,If + O6\,V    & Y & N & -- \\
MT91\,421     & O9\,V + B9\,V-A0\,V & Y & N & -- \\
Cyg\,OB2 \#7  & O3\,If            & Y & N & -- \\
MT91\,516     & O5.5\,V           & Y & N & -- \\
MT91\,534     & O7.5\,V           & Y & Y & 10960\\
Cyg\,OB2 \#11 & O5\,If + B0\,V    & Y & N & -- \\
Cyg\,OB2 \#75 & O9\,V             & Y & N & -- \\ 
Cyg\,OB2 \#73 & O8\,III + O8\,III & Y & N & -- \\
MT91\,771     & O7\,V + O9\,V     & Y & N & -- \\
\hline
\multicolumn{5}{|c|}{B-stars}\\
\hline
MT91\,103     & B1\,V + B2\,V      & Y & & \\  
MT91\,213     & B0\,V              & Y & & \\  
MT91\,336     & B3\,III           & Y & & \\  
MT91\,620     & B0\,V             & Y & & \\  
MT91\,646     & B1.5\,V            & Y & & \\  
MT91\,759     & B1\,V              & Y & & \\  
\hline
\multicolumn{5}{|c|}{WR-stars}\\
\hline
WR\,145       & WN7o/CE + O7\,V((f)) & - & Y & 10969 \\  
\hline
\end{tabular}
\end{center}
\end{table*}

In all our models, the ISM column density was set to $1.32 \times 10^{22}$\,cm$^{-2}$. For the non-solar abundance model ({\tt phabs*vphabs*vapec(2T)}), the best fit ($\chi_{\nu}^2 = 1.32$ for 176 degrees of freedom) is achieved with $N_{\rm wind} = (6.2^{+2.1}_{-1.7}) \times 10^{18}$\,cm$^{-2}$, $kT_1 = 0.36^{+.09}_{-.08}$ and $kT_2 = 2.10^{+.25}_{-.22}$\,keV. As for WR\,145, the low value of $N_{\rm wind}$ indicates that the bulk of the X-ray emission probably arises from either the wind-wind interaction zone or the O8 companion. We have thus repeated the fitting process adopting solar abundances for both the emitting and absorbing gas (see Fig.\,\ref{WRspec}). As for WR\,145, this model provides a somewhat better adjustment ($\chi_{\nu}^2 = 1.23$ for 176 degrees of freedom), and the parameters are $N_{\rm wind} = (0.46^{+.20}_{-.17}) \times 10^{22}$\,cm$^{-2}$, $kT_1 = 0.36^{+.11}_{-.08}$ and $kT_2 = 2.1^{+.4}_{-.3}$\,keV. The observed and ISM corrected X-ray fluxes are equal to $3.0 \times 10^{-13}$ and $13.6 \times 10^{-13}$\,erg\,cm$^{-2}$\,s$^{-1}$ respectively. Adopting an O8\,I classification for the companion, we find that the dereddened flux agrees extremely well with the level expected from the sole O-star via our $\log{\rm L_{\rm X}/{\rm L}_{\rm bol}}$ relation. It thus seems quite likely that part of the X-ray emission of WR\,146 arises from the O-star companion. 
To further investigate the temporal dependence, we have folded our best-fit (solar composition) model through the response matrices of the {\it EINSTEIN}-IPC and {\it ROSAT}-PSPC-C instruments. The predicted count rates are $(6.7 \pm 1.1) \times 10^{-3}$\,cts\,s$^{-1}$ for the IPC, and $(7.5 \pm 3.1) \times 10^{-3}$\,cts\,s$^{-1}$ for the PSPC-C. The agreement with the {\it EINSTEIN} data is very good, whilst the {\it ROSAT} All Sky Survey observations yield a count rate a factor two lower than predicted here. Given the uncertainties (mainly reflecting the short integration time of the RASS on this source), this difference is however not highly significant. We thus conclude that there is currently no firm evidence for long-term variations in the X-ray flux of WR\,146.

\section{Inter-pointing X-ray variability}
In single massive stars that emit X-rays through the LDI mechanism, a large number of pockets of shock-heated X-ray plasma are expected to be scattered throughout the wind volume. The resulting X-ray emission is usually not seen to vary \citep{zetaPup2}. However, considerable variability can arise either as a result of rotational modulation in single stars with magnetically confined winds \citep{BabelMontmerle}, or in massive binary systems that host a wind interaction zone \citep{PS}. The tiling strategy employed for the survey implies that most stars are observed typically four times over a six-week period. In addition, the combination of the legacy survey with older data \citep{WD} allows us to check for long-term variability. We can do this using either the exposure-corrected count rates or the fluxes inferred from spectral fits. 
\subsection{Count rates}
As a first step towards a quantification of the inter-pointing X-ray variability of massive stars in Cyg\,OB2, we consider the photon fluxes corrected for averaged observatory response over the [0.5,8]\,keV energy band \citep[see Eq.\,8 of][]{Broos}. For each source, we have performed a $\chi^2$ variability test on the count rates recorded during the various observations. Out of the 108 detected sources (51 O-stars, 54 B-stars, 3 WR stars), 23 are found to be variable at the 1\% significance level. These 23 objects include the brightest stars (Cyg\,OB2 \#5, 8a, 9) which suffer from pile-up in the ACIS observations. The photometry of these objects could thus also be affected and we leave them aside in the following. The results are given in Table\,\ref{variab}. We further identify three stars which show short-term variability (during a single observation) at a significance level of $\leq 1$\% according to a Kolmogorov-Smirnov test. 


\subsection{Fluxes}
Based on our spectral fits, we have also investigated the variability of the X-ray flux of the brightest sources that were observed several times. We focus on the fluxes as they are generally much better constrained than other spectral model parameters, such as plasma temperatures and wind column densities, which are affected by parameter correlations. Yet, we caution that, because of the larger error bars on the fluxes compared to the count rates, some objects found to be variable in the previous section are found not to be significantly variable in terms of their fluxes.

\begin{figure*}[thb]
\begin{minipage}{8cm}
\begin{center}
\resizebox{8cm}{!}{\includegraphics{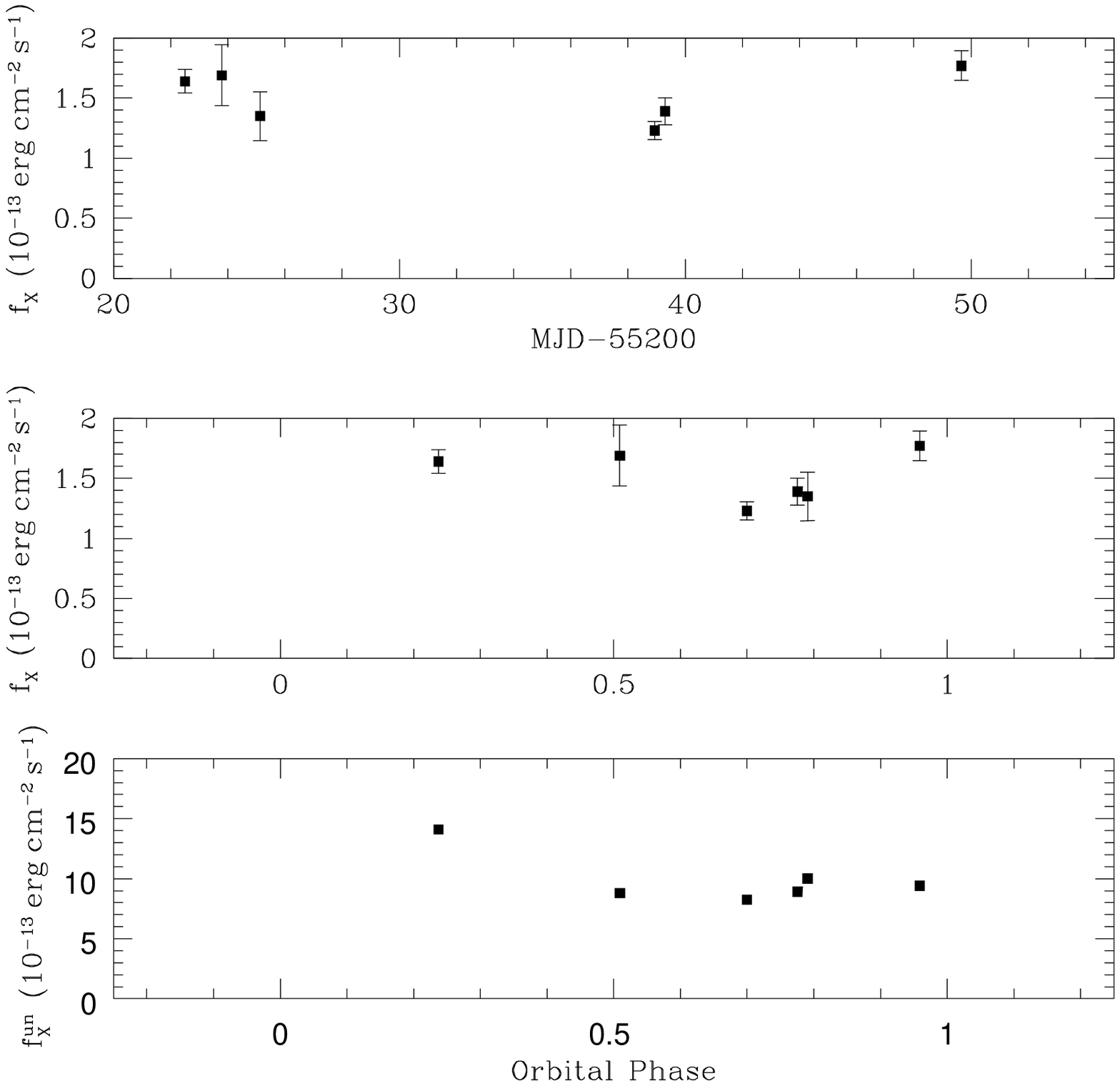}}
\end{center}
\end{minipage}
\hfill
\begin{minipage}{8cm}
\begin{center}
\resizebox{8cm}{!}{\includegraphics{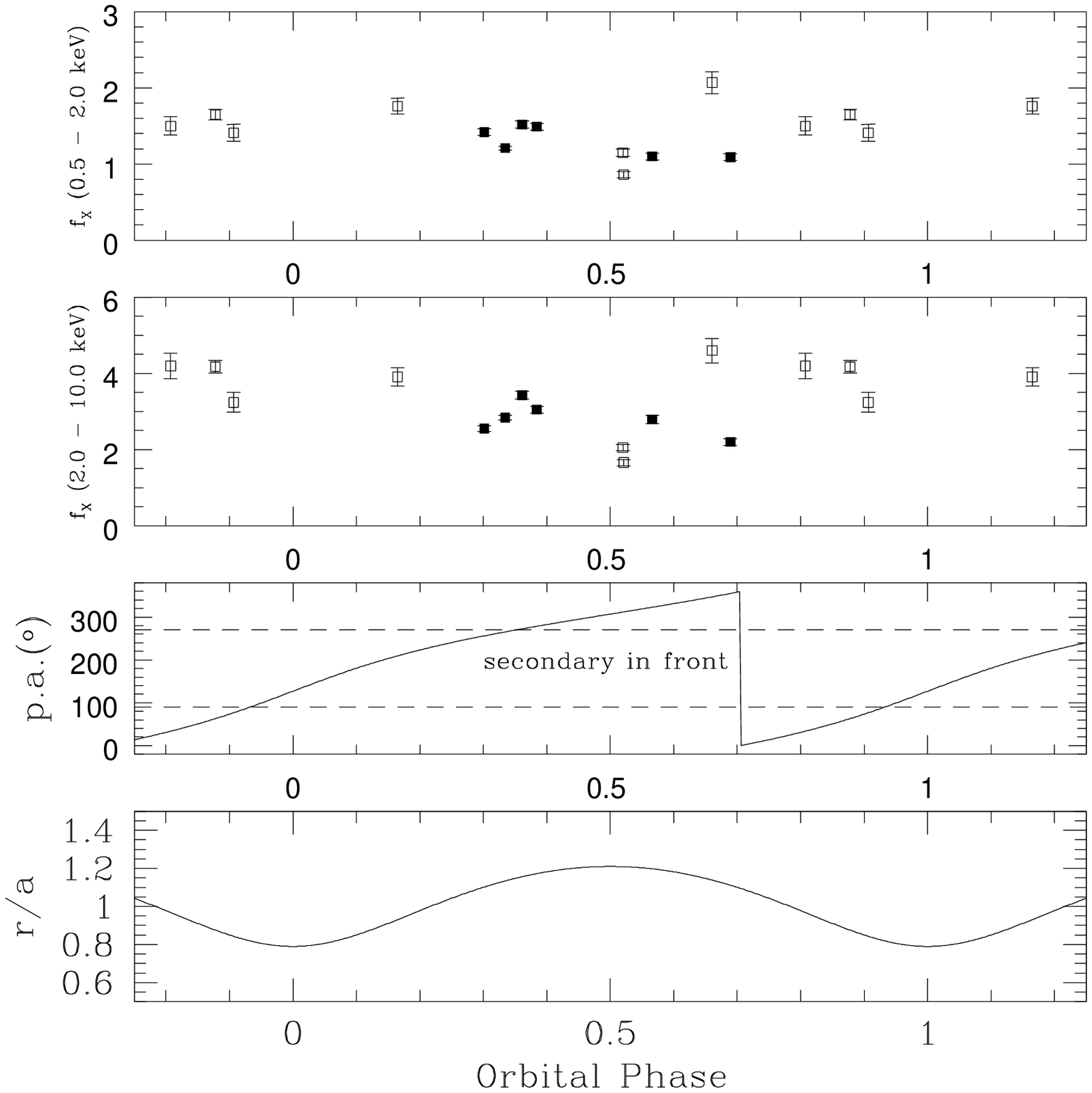}}
\end{center}
\end{minipage}
\caption{Left: epoch and phase-dependence of the observed and absorption-corrected X-ray fluxes of Cyg\,OB2 \#3. The orbital phases were computed using the ephemerides from \citet{Kiminki08}. Right, from top to bottom: observed X-ray flux in the soft (0.5 -- 2\,keV) and hard (2 -- 10\,keV) bands (in units $10^{-13}$\,erg\,cm$^{-2}$\,s$^{-1}$), position angle (defined as p.a.\ = 0$^{\circ}$ at conjunction with the primary in front) and relative orbital separation of CPR2002\,A11. The filled and open symbols stand respectively for the ACIS and EPIC data.\label{Schulte3}}
\end{figure*}

\subsubsection{Known binary systems \label{CWBs}}
In colliding wind binary systems, one can expect phase-locked variability of the X-ray spectrum as a result of a changing column density along the line of sight towards the wind-wind interaction zone as the stars move around their common center of mass. In eccentric systems, additional variations are expected as the physical properties of the wind-wind interaction change with the changing separation. If the wind interaction zone is in the adiabatic regime \citep{SBP}, one expects to observe an orbital modulation of the X-ray flux as $1/d$ where $d$ is the instantaneous separation between the stars, as is actually observed in the long-period system Cyg\,OB2 \#9 \citep{Naze9}. 

\paragraph{Cyg\,OB2 \#3}
According to \citet{Kiminki08}, Cyg\,OB2 \#3 is an O6\,IV + O9\,III eclipsing binary system with an orbital period of 4.7464\,days and an almost circular orbit ($e = 0.07$). The ACIS spectra from individual observations can be fitted using a single temperature plasma model. The intrinsic X-ray spectrum appears rather soft, with a mean\footnote{The uncertainties quoted correspond to the dispersion about the mean.} $kT$ of  $(0.83 \pm 0.11)$\,keV and moderate absorption by wind material ($N_{\rm wind} = (0.20 \pm 0.10) \times 10^{22}$\,cm$^{-2}$).   

Our observations of this star (collected over a total time span of 27 days) sample more than half of the orbital cycle (see Fig.\,\ref{Schulte3}). The observed flux varies by 14\% (standard deviation about the mean)\footnote{Pile-up should not be a critical issue for this source, as we estimate corrections on the observed X-ray fluxes of at most 3\%.}. This is compatible with the estimated relative errors on the fluxes of individual pointings, which are between 6 and 15\%.  The ISM corrected fluxes show larger variability (22\%), but are also subject to larger uncertainties (partly due to the degeneracy between plasma temperature and wind column). In summary, we conclude that there is no clear evidence for significant orbital modulation of the X-ray flux of Cyg\,OB2 \#3. 
\paragraph{CPR2002\,A11 = MT91\,267}
The SB1 status of this O7\,I star was recently reported by \citet{Kobulnicky} who derived an orbital period of 15.511\,days and an eccentricity of 0.21. From {\it XMM-Newton} observations, its X-ray emission was found to be variable with flux variations by more than a factor two \citep{Rauw}. Most individual ACIS spectra require a two temperature plasma to achieve a decent fit and for consistency, we have repeated the fitting of the EPIC data with the same model as for the ACIS spectra.

We have computed the orbital phases of the observations using the ephemerides of \citet{Kobulnicky}. Combining the {\it XMM-Newton} and {\it Chandra} data, we have an almost complete coverage of the orbital cycle, except near phase 0.0 (periastron). To better constrain the origin of the variability, we distinguish the observed fluxes over two energy domains: a soft band (0.5 -- 2\,keV) and a hard band (2 -- 10\,keV). The results are shown in Fig.\,\ref{Schulte3}, along with a plot of the relative orbital separation and the position angle (defined as p.a.\ = $0^{\circ}$ when the O7\,I primary star is in front). Although there are some hints that the flux in the hard band is lower near phase 0.5, this needs confirmation. Indeed, the EPIC data show a larger amplitude of variability than the ACIS data, and observations taken at similar orbital phases sometimes display rather different fluxes. Whilst there could be some remaining discrepancies between the calibration of the EPIC and ACIS responses (see Sect.\ \ref{global}), it seems unlikely that they could account for the observed differences. The ACIS spectra could suffer from pile-up, but based on the EPIC spectra, we have estimated that the ACIS pile-up fraction should be less than 9\%. Including the {\tt pileup} command in the fit, leads to slightly different fluxes (typical differences of 7\%), but does not change the general appearance of the plot in Fig.\,\ref{Schulte3}. 

\paragraph{Cyg\,OB2 \#22 \label{sec22}}
Cyg\,OB2 \#22 is a multiple system consisting of an O3\,I component (star A) and an O6\,V star \citep[star B,][]{Walborn1,Mason} separated by about 1.5\,arcsec. The O6\,V component is itself a double system with a separation of 0.2\,arcsec and a magnitude difference of 2.34 in the $z$ filter \citep{Sota}, and is furthermore found to be an SB1 binary with a period near 35 days (Kobulnicky et al., in preparation). Components A and B are resolved with {\it Chandra} when the star falls on-axis (ObsIDs 4511 and 10956), but are highly confused otherwise. Although the {\tt acisextract} routine attempts to extract sources A and B as a pair on all our observations, those data taken at large off-axis angles, must be considered with caution\footnote{In the X-ray to bolometric luminosity relations, we have considered the sum of components A and B.}.

Based on six {\it XMM-Newton} observations, \citet{Rauw} reported on X-ray variability of this system with the combined X-ray flux of A + B varying by a factor 1.75 within ten days, a time-scale potentially related to the newly found orbital period of Cyg\,OB2 \#22 B. Including all the {\it XMM-Newton} data, flux variability by a factor 2.1 was found.  
  
\begin{figure}[h]
\begin{center}
\resizebox{8cm}{!}{\includegraphics{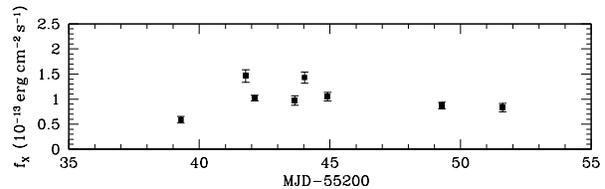}}
\end{center}
\caption{Combined X-ray flux of Cyg\,OB2 \#22 components A and B as a function of time during the ACIS survey program.\label{Schulte22}}
\end{figure}

To allow comparison with the EPIC data and because of the above described difficulties with the source extraction, we have summed the fluxes from the ACIS data of components A and B (see Fig.\,\ref{Schulte22}). Although this result must be taken with caution, it seems that the ACIS data indeed support the existence of flux variations on time-scales of a few days. 

In principle, we should be able to combine the fluxes from the ACIS and EPIC data to perform a Fourier analysis. However, the difficulties with the source extraction described above could impact on the result. Moreover, comparing the fluxes found with {\it XMM-Newton} and {\it Chandra}, we notice that the former are systematically larger than the latter by a factor 1.5 -- 2.0. 
 
Therefore, whilst it seems that the X-ray flux of Cyg\,OB2 \#22 A+B is variable on a rather short term, the current data do not allow us to establish the exact value of this time-scale. 

\begin{figure}[h]
\begin{center}
\resizebox{8cm}{!}{\includegraphics{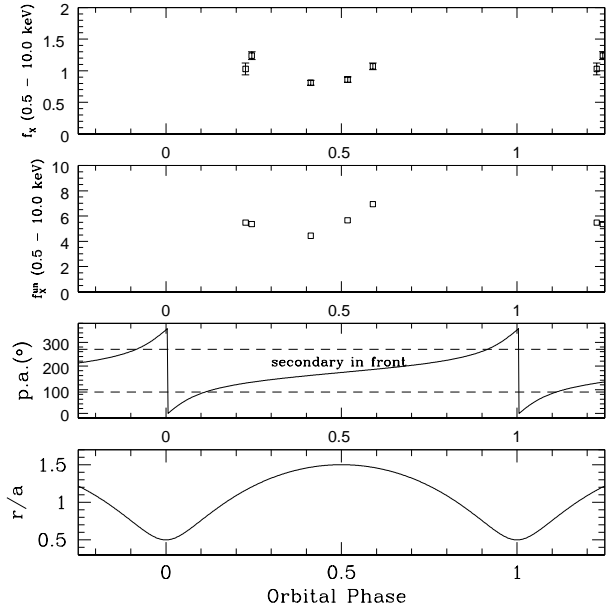}}
\end{center}
\caption{From top to bottom: observed and ISM-absorption corrected X-ray fluxes (in units $10^{-13}$\,erg\,cm$^{-2}$\,s$^{-1}$), position angle and relative orbital separation of Cyg\,OB2 \#11. \label{Schulte11}}
\end{figure}

\paragraph{Cyg\,OB2 \#11}
Cyg\,OB2 \#11 (O5\,I) was reported as an eccentric ($e = 0.50$) SB1 binary with a period of 72.4\,days by \citet{Kobulnicky}. According to the ephemerides provided by the latter authors, our observations span a bit more than one third of the cycle roughly centered on phase $\phi = 0.4$ (periastron passage corresponding to $\phi = 0.0$; see Fig.\,\ref{Schulte11}). The observed flux varies by 17\% (standard deviation about the mean), which is significant, given that the typical relative errors on individual data points are of order 5\%. Cyg\,OB2 \#11 is thus a good candidate for a phase-locked variation of the X-ray flux due to wind-wind interactions, and it would be interesting to collect observations near periastron passage. If the wind interaction zone is in the adiabatic regime, we would then expect the X-ray flux to be about three times larger than measured during the present campaign. 

\paragraph{MT91\,771}
MT91\,771 is an O7\,V + O9\,V binary with an almost circular orbit ($e = 0.05 \pm 0.03$) and a short period of 2.82\,days \citep{Kiminki12}. Our observations reveal little evidence for variability. The standard deviation around the mean observed flux amounts to 9\% (with relative errors on individual fluxes of 6\%). 

\subsubsection{Binary candidates}
MT91\,138 and Cyg\,OB2 \#8c are radial velocity variable stars listed respectively as SB1 and SB1? by \citet{Kiminki}, although no orbital solution is available for any of these stars. 

Our data include four observations of MT91\,138 over 26 days. The observed X-ray flux of this star remains constant to within 6\%, which is well below the typical relative uncertainty of 11\%. 

For Cyg\,OB2 \#8c, we have five observations at hand, four of them are from the survey and span two days. The fluxes of the source seem to vary at the 20\% level (typical uncertainties being 7\%). One must be careful though with this source, as it falls very close to the bright Cyg\,OB2 \#8a which could contaminate its spectrum or the background spectrum especially for observations taken at relatively large off-axis angle. 

We thus conclude that there is currently no clear indication for flux variability of these sources.

\subsubsection{Probably single stars}
CPR2002\,A20 was observed twice, separated by 2.5\,days. No significant variability is found in the observed fluxes of this source. 

MT91\,534 was observed five times, at first during the original Cyg\,OB2 {\it Chandra} observation and six years later in the course of the survey. Typical errors on observed fluxes of individual pointings range between 5 and 15\%. Except for one pointing (ObsID 10960), the fluxes are relatively constant: the standard deviation about the mean ($2.6 \times 10^{-14}$\,erg\,cm$^{-2}$\,s$^{-1}$) of the observed fluxes amounts to 6\% of the mean flux\footnote{This star was slightly fainter than this mean level at the time of the {\it XMM-Newton} observations (see its position in Fig.\ \ref{ACISpn}).}. ObsID 10960 is a clear outlier: the observed flux ($3.19 \times 10^{-13}$\,erg\,cm$^{-2}$\,s$^{-1}$) is a factor 12 larger during this observation than during any other pointing and the star shows clear intra-pointing variability (see Table\,\ref{variab}). Furthermore, the plasma temperature is much higher than on average (5.0 versus 1.7\,keV). These properties are reminiscent of flares in low-mass pre-main sequence stars. This flare could potentially reveal an otherwise undetectable low-mass companion near the O-star. In this context, we note that the ISM-corrected flux at ObsID 10960 ($5.14 \times 10^{-13}$\,erg\,cm$^{-2}$\,s$^{-1}$) corresponds to an X-ray luminosity of $1.2 \times 10^{32}$\,erg\,s$^{-1}$. Such values are certainly not unusual for flaring late-type pre-main sequence stars \citep[e.g.][]{Wolk}.

Cyg\,OB2 \#7 was observed five times with {\it Chandra} (during the original Cyg\,OB2 observation and six years later in the course of the survey, four times within 2 days), as well as at seven epochs with {\it XMM-Newton}. Based on six of the seven {\it XMM-Newton} observations, \citet{Rauw} concluded that this star was constant to within 10\% in the {\it XMM-Newton} data. The seventh {\it XMM-Newton} observation yields a somewhat lower flux which deviates by 22\% from the mean of the previous six spectra. For the ACIS spectra, we find a dispersion about the mean of 7\%, supporting the idea that the source is constant at least on relatively short time-scales. 
The EPIC data were well fitted using a single plasma component. The ACIS spectra are usually better fitted by including a second plasma component, although there is a degeneracy between the plasma temperature and the wind column density, and the second (higher) plasma temperature is usually only very poorly constrained. Comparing the single plasma component fits, we find an observed flux of $(1.70 \pm 0.18) \times 10^{-13}$\,erg\,cm$^{-2}$\,s$^{-1}$ for the {\it XMM-Newton} data , whilst the ACIS spectra yield $(1.15 \pm 0.09) \times 10^{-13}$\,erg\,cm$^{-2}$\,s$^{-1}$, i.e.\ a difference of 50\%. A priori, pile-up should not be an issue for this object (estimated pile-up fractions are 2 -- 4\%). We have nevertheless also performed a 2-T fit of the ACIS spectra including the {\tt pileup} command. However, the observed flux remains at a low level of $(1.22 \pm 0.10) \times 10^{-13}$\,erg\,cm$^{-2}$\,s$^{-1}$. Contamination of the EPIC data by nearby weak point sources is the most likely explanation of the difference between the ACIS and EPIC fluxes.

Cyg\,OB2 \#8b was observed at the same five epochs as Cyg\,OB2 \#7. Typical errors on the determination of the observed fluxes are between 6 and 16\%. The standard deviation about the mean of the observed flux is 24\%. 
One has to bear in mind that this source is located in a complex region (see our remarks on Cyg\,OB2 \#8c).

\begin{figure}[h]
\begin{center}
\resizebox{8cm}{!}{\includegraphics{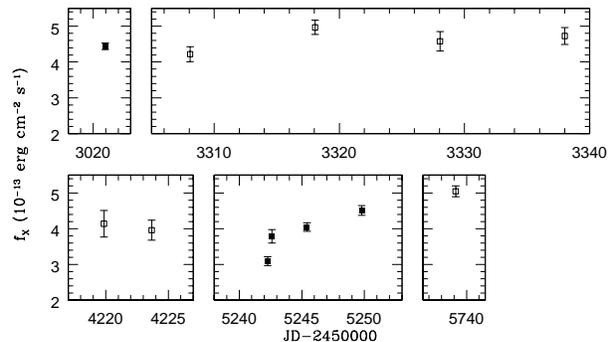}}
\end{center}
\caption{Observed flux of MT91\,516 as a function of time. The open symbols stand for {\it XMM-Newton} data from \citet{Rauw} whilst the filled symbols indicate fluxes inferred from the ACIS data presented here.\label{MT516}}
\end{figure}

MT91\,516 was also observed five times. As this source is at the limit of a moderate pile-up, we have used the {\tt pileup} command in the fits. The resulting observed fluxes are found to vary by 36\% (peak to peak). This is slightly larger than the 22\% flux variability found by \citet{Rauw} in the {\it XMM-Newton} data of this object\footnote{The seventh {\it XMM-Newton} observation \citep{Naze9} yields an observed flux of $5.07 \times 10^{-13}$\,erg\,cm$^{-2}$\,s$^{-1}$, slightly above the highest value of the other six observations discussed by \citet{Rauw}.}. On average, the ACIS fluxes are somewhat lower (12\%) than the EPIC results, although, in the case of cyclic variability, this could also be due to a difference in the sampling of the variability cycle. Variability could hint at binarity with a likely time scale of order a few weeks (see Fig.\ \ref{MT516}), although \citet{Kiminki} reported a probability of 11.5\% that the radial velocity variations of this star be spurious, and from 12 radial velocity measurements, Kobulnicky et al.\ (in preparation) found no significant variations.  

\begin{figure}[h]
\begin{center}
\resizebox{8cm}{!}{\includegraphics{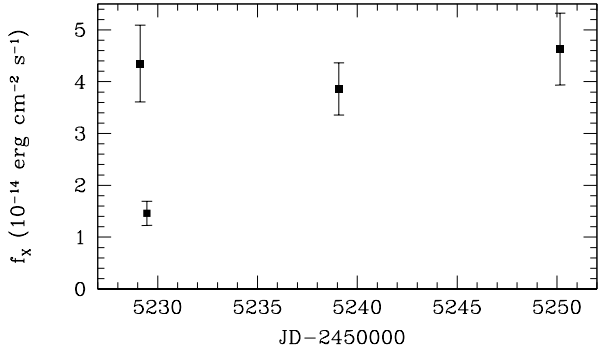}}
\end{center}
\caption{Observed flux of MT91\,213 as a function of time during the survey.\label{MT213}}
\end{figure}

MT91\,213 (B0\,V) is the only B-star\footnote{This is actually a Be star with variable emission lines (Kobulnicky et al.\ in preparation).} that was found to be variable and has a sufficient number of counts nearly each time it was observed to perform a spectral fit. The variations of the flux during the survey are shown in Fig.\,\ref{MT213}. There are very rapid variations (by a factor 3) between ObsIDs 10944 and 10945, i.e.\ within less than 8 hours. One may wonder whether the flux of ObsID 10945 is reliable, as it is the only data point that strongly deviates. However, an older observation (ObsID 4501) actually indicates a very similar, even somewhat lower, flux of $1.21 \times 10^{-14}$\,erg\,cm$^{-2}$\,s$^{-1}$. Thus, we conclude that the flux of MT91\,213 indeed varies by at least a factor three and that these variations can occur on short time scales. Flares from an unseen low-mass companion are unlikely to explain this behavior, as the star seems to spend more time at a roughly constant high flux level, unlike what is seen in flaring stars.     

\section{Summary and conclusions}
In this paper, we have analyzed one of the richest samples of X-ray data of OB stars belonging to a single association. These data have shed new light on the X-ray properties of massive stars. 

We have shown that O-stars in Cyg\,OB2 follow a well-defined scaling relation between their X-ray and bolometric luminosities with $\log{\frac{{\rm L}_{\rm X}}{{\rm L}_{\rm bol}}} = -7.2 \pm 0.2$. This relation is in excellent agreement with the one previously derived from {\it Chandra} observations of the Carina Nebula, suggesting that any environmental effect on this relation should be quite small. 
Our investigation indicates that, owing to its narrow PSF, {\it Chandra} is the best mission to evaluate the X-ray emission of moderately bright and faint massive stars in crowded environments, whilst {\it XMM-Newton} is better suited to study the massive stars at the X-ray brighter end.

Except for the brightest O-star binaries, that we have not studied here, we do not find a general X-ray overluminosity due to colliding winds in O-star binaries, neither do we find a clear phase-locked variability in most of them. For O-star binary systems with short orbital periods, there is some tentative trend for an increase of the X-ray overluminosity with wind kinetic power, although this result clearly calls for confirmation. 

B-type stars do not show a clear relationship between their X-ray and bolometric luminosity, suggesting that their X-ray emission might come, at least for some of them, from a low-mass companion. 

Finally, out of the three WC stars in Cyg\,OB2, probably only one (WR\,144) is itself responsible for the observed level of X-ray emission, $\log{\frac{{\rm L}_{\rm X}}{{\rm L}_{\rm bol}}} = -8.8 \pm 0.2$. The X-ray emission of the other two Wolf-Rayet stars in Cyg\,OB2 can be accounted for by the emission of their O-type companion as well as a moderate contribution from a wind-wind interaction zone.


\acknowledgments

The Li\`ege team acknowledges support from Belspo through an XMM PRODEX contract, from the FRS/FNRS and from an ARC grant for Concerted Research Actions, financed by the Federation Wallonia-Brussels. MG and NJW acknowledge support from {\it Chandra} grant GO0-11040X. NJW was also supported by a Royal Astronomical Society research fellowship. JJD was supported by NASA contract NAS8-03060 to the {\it Chandra} X-ray Center and thanks the director, H.\ Tananbaum, and science team for continuing advice and support.



{\it Facilities:} \facility{Chandra (ACIS)}, \facility{XMM-Newton (EPIC)}.



\appendix
\section{More details on the ${\rm L}_{\rm X}/{\rm L}_{\rm bol}$ relation \label{annexe}}
\citet{AC} argued that at least for evolved OB stars (luminosity classes I-III) in Cyg\,OB2, a lower scatter is achieved for power-law relations instead of simple scaling laws.

We have therefore tested a power law relation on our full data set, where L$_{\rm X}$ is allowed to scale with some power of L$_{\rm bol}$. The best-fit power-law relation (for equal weights of all 40 data points) becomes
$$\log{{\rm f}_{\rm X}} = (1.24 \pm 0.12)\,\log{{\rm f}_{\rm bol}} -(5.85 \pm 0.69)$$
Adopting instead a weighting according to the square root of the number of counts in the spectrum yields
$$\log{{\rm f}_{\rm X}} = (1.20 \pm 0.09)\,\log{{\rm f}_{\rm bol}} -(6.02 \pm 0.48)$$
In the same way, weighting the data according to the estimated errors on the fluxes leads to
$$\log{{\rm f}_{\rm X}} = (1.19 \pm 0.08)\,\log{{\rm f}_{\rm bol}} -(6.14 \pm 0.46)$$

We have then considered the 13 O-type stars of our sample with luminosity class I-III. These objects have 
$$\log{{\rm L}_{\rm X}/{\rm L}_{\rm bol}} = -7.18 \pm 0.19$$
(unweighted), and 
$$\log{{\rm L}_{\rm X}/{\rm L}_{\rm bol}} = -7.15 \pm 0.19$$
(weighted; same relation for both types of weights). A power-law relation for the unweighted and weighted samples of giants and supergiants yields respectively 
$$\log{{\rm f}_{\rm X}} = (1.04 \pm 0.09)\,\log{{\rm f}_{\rm bol}} -(6.97 \pm 0.51)$$ 
and 
$$\log{{\rm f}_{\rm X}} = (1.16 \pm 0.13)\,\log{{\rm f}_{\rm bol}} -(6.33 \pm 0.65)$$ 
for weighting according to the square root of the total number of counts, and
$$\log{{\rm f}_{\rm X}} = (1.14 \pm 0.12)\,\log{{\rm f}_{\rm bol}} -(6.41 \pm 0.62)$$ 
for weighting according to the estimated errors on ISM-corrected X-ray fluxes.

We see that, independently of the weighting, and of whether or not we restrict ourselves to the giants and supergiants, the exponent of L$_{\rm bol}$ deviates by less than 2\,$\sigma$ from unity (i.e.\ from a simple scaling law). Our data thus do not support the need of a power-law relation, suggesting that a scaling law is sufficient to describe the dependence of L$_{\rm X}$ on L$_{\rm bol}$ for O-stars of all luminosity classes.

Restricting ourselves to the 23 presumably single stars of the cleaned sample, we obtain the scaling relation
$$\log{{\rm L}_{\rm X}/{\rm L}_{\rm bol}} = -7.27 \pm 0.21$$
$$\log{{\rm L}_{\rm X}/{\rm L}_{\rm bol}} = -7.22 \pm 0.21$$
and 
$$\log{{\rm L}_{\rm X}/{\rm L}_{\rm bol}} = -7.23 \pm 0.19$$
for the unweighted data, the data weighted according to the number of counts, and the data weighted according to the estimated errors respectively.
The corresponding power law relations are
$$\log{{\rm f}_{\rm X}} = (1.04 \pm 0.20)\,\log{{\rm f}_{\rm bol}} -(7.06 \pm 1.16)$$
$$\log{{\rm f}_{\rm X}} = (1.13 \pm 0.27)\,\log{{\rm f}_{\rm bol}} -(6.49 \pm 1.60)$$
and 
$$\log{{\rm f}_{\rm X}} = (1.09 \pm 0.27)\,\log{{\rm f}_{\rm bol}} -(6.72 \pm 1.58)$$
Within the error bars, there is no significant difference between the relations obtained from the full sample and those for presumably single stars only, even though including binaries leads to systematically higher average $\log{{\rm L}_{\rm X}/{\rm L}_{\rm bol}}$ values \citep[as was also found e.g.\ in the Carina Nebula,][]{Carina}.

\section{Spectral fits}
\begin{sidewaystable}
\caption{Spectral fits of O-type stars \label{fitsO}}
\tiny
\begin{tabular}{l c c c c c c c c c c c c c c}
\hline
\multicolumn{1}{c}{Star} & Spectral type & $\alpha$ & $\delta$ & Total counts & $N_{\rm H}^{ISM}$ & $N_{\rm wind}$ & $kT_1$ & norm$_1$ & kT$_2$ & norm$_2$ & $f_{\rm X}^{\rm obs}$ & $f_{\rm X}$ & $\chi^2_{\nu}$ & $\nu$ \\
& & (J2000.0) & (J2000.0) & & ($10^{22}$\,cm$^{-2}$) & ($10^{22}$\,cm$^{-2}$) & (keV) & & (keV) & & \multicolumn{2}{c}{($10^{-13}$\,erg\,cm$^{-2}$\,s$^{-1}$)} & & \\
\hline
MT91\,5 & O6\,V((f)) & 203039.82 & +413650.5 & 15.1 & 1.11 & $0.12^{+2.26}_{-0.12}$ & $0.37^{+.66}_{-.31}$ & $8.0 \times 10^{-5}$ & -- & -- & 0.05 & 1.04 & 0.49 & 3 \\
CPR2002\,A26 & O9.5\,V & 203057.61 & +410956.6 & 13.0 & 1.26 & 0.0 & $0.14^{+.14}_{-.05}$ & $5.2 \times 10^{-3}$ & -- & -- & 0.04 & 32.5 & 0.05 & 1 \\
Cyg\,OB2 \#1 & O8\,V & 203110.53 & +413153.4 & 87.0 & 0.99 & $0.16^{+.31}_{-.16}$ & $0.49^{+.19}_{-.19}$ & $8.0 \times 10^{-5}$ & -- & -- & 0.10 & 1.16 & 0.53 & 14 \\
MT91\,70 & O9\,V & 203118.33 & +412121.8 & 29.0 & 1.37 & $0.52^{+2.75}_{-.52}$ & $0.16^{+.28}_{-.11}$ & $2.14 \times 10^{-3}$ & -- & -- & 0.02 & 0.73 & 0.79 & 8 \\
CPR2002\,A15 & O7\,I & 203136.91 & +405909.4 & 111.6 & 1.46 & 0.0 & $0.60^{+.13}_{-.14}$ & $9.9 \times 10^{-5}$ &  -- & -- & 0.13 & 2.93 & 1.06 & 20 \\
Cyg\,OB2 \#3 & O6\,IV + O9\,III & 203137.50 & +411321.1 & 1445.2 & 1.11 & $0.51^{+.17}_{-.17}$ & $0.24^{+.14}_{-.06}$ & $5.55 \times 10^{-3}$ & $0.92^{+.12}_{-.10}$ & $4.56 \times 10^{-4}$ &  1.52 &  10.83 & 1.09 & 125\\
MT91\,138 & O8.5\,I & 203145.40 & +411826.9 & 353.2 & 1.31 & 0.0 & $0.59^{+.09}_{-.11}$ & $1.51 \times 10^{-4}$ & -- & -- & 0.22 & 4.45 & 1.06 & 53 \\
MT91\,140 & O9.5\,I & 203145.97 & +411727.0 & 92.7 & 0.31 & 0.0 & $0.28^{+.08}_{-.10}$ & $1.9 \times 10^{-5}$ & -- & -- & 0.08 & 0.35 & 0.62 & 16 \\
Cyg\,OB2 \#20 & O9\,III & 203149.66 & +412826.3 & 56.9 & 0.79 & 0.0 & $0.27^{+.19}_{-.14}$ & $3.6 \times 10^{-5}$ & -- & -- & 0.02 & 0.40 & 0.71 & 12 \\
CP2012\,E45 & O7\,V & 203159.63 & +411450.2 & 87.6 & 1.10 & 0.0 & $0.91^{+.19}_{-.37}$ & $2.23 \times 10^{-5}$ & -- & -- & 0.08 & 0.64 & 0.54 & 14 \\ 
Cyg\,OB2 \#4 & O7\,III & 203213.84 & +412711.4 & 296.9 & 0.84 & $0.26^{+.20}_{-.18}$ & $0.37^{+.14}_{-.10}$ & $2.1 \times 10^{-4}$ & -- & -- & 0.16 & 1.51 & 0.74 & 42 \\
Cyg\,OB2 \#14 & O9\,V & 203216.57 & +412535.7 & 104.5 & 0.88 & 0.0 & $0.60^{+.19}_{-.42}$ & $1.3 \times 10^{-5}$ & -- & -- & 0.04 & 0.38 & 0.83 & 23 \\
Cyg\,OB2 \#15 & O8\,V & 203227.66 & +412622.1 & 41.1 & 0.85 & $0.10^{+.36}_{-.10}$ & $0.54^{+.92}_{-.37}$ & $4.3 \times 10^{-5}$ & -- & -- & 0.09 & 0.85 & 0.68 & 4 \\
CPR2002\,A11 & O7.5\,III & 203231.53 & +411408.1 & 7482.8 & 1.43 & $0.23^{+.05}_{-.05}$ & $0.84^{+.10}_{-.06}$ & $8.19 \times 10^{-4}$ & $2.31^{+.48}_{-.18}$ & $4.46 \times 10^{-4}$ & 4.20 & 17.10 & 1.11 & 301 \\
CPR2002\,A38 & O8\,V & 203234.87 & +405617.0 & 117.4 & 1.08 & 0.0 & $0.85^{+.23}_{-.31}$ & $1.88 \times 10^{-5}$ & -- & -- & 0.06 & 0.56 & 0.83 & 67 \\
Cyg\,OB2 \#16 & O8\,V & 203238.55 & +412513.6 & 286.8 & 0.86 & 0.0 & $0.68^{+.07}_{-.09}$ & $3.3 \times 10^{-5}$ & -- & -- & 0.11 & 1.01 & 1.29 & 48 \\
Cyg\,OB2 \#6 & O8\,V & 203245.44 & +412537.6 & 264.3 & 0.88 & 0.0 & $0.49^{+.14}_{-.09}$ & $6.02 \times 10^{-5}$ & -- & -- & 0.12 & 1.63 & 1.09 & 43 \\
Cyg\,OB2 \#17 & O8.5\,V & 203250.01 & +412344.5 & 171.0 & 0.94 & 0.0 & $0.98^{+.14}_{-.12}$ & $1.77 \times 10^{-5}$ & -- & -- & 0.08 & 0.47 & 0.66 & 33 \\
MT91\,376 & O8\,V & 203259.16 & +412425.3 & 96.9 & 0.94 & $0.39^{+.39}_{-.27}$ & $0.22^{+.07}_{-.07}$ & $5.67 \times 10^{-4}$ & -- & -- & 0.05 & 0.97 & 0.79 & 15 \\
MT91\,390 & O8\,V & 203302.92 & +411743.1 & 139.9 & 1.31 & 0.0 & $0.98^{+.18}_{-.19}$ & $2.18 \times 10^{-5}$ & -- & -- &  0.07 & 0.58 & 0.74 & 22 \\
CPR2002\,A20 & O8\,II & 203302.93 & +404725.2 & 3190.7 & 1.28 & $0.76^{+.14}_{-.14}$ &  $0.19^{+.11}_{-.04}$ & $4.61 \times 10^{-2}$ & $1.18^{+.09}_{-.06}$ &  $2.00 \times 10^{-3}$ &  5.85 & 24.80 & 0.88 & 207 \\
Cyg\,OB2 \#22 & O3\,If + O6\,V & 203308.77 & +411318.7 & 1522.0 & 1.35 & $0.46^{+.11}_{-.10}$ & $0.45^{+.05}_{-.05}$ & $1.58 \times 10^{-3}$ & $> 2.83$ & $1.97 \times 10^{-5}$ & 1.04 & 7.85 & 1.26 & 118 \\
MT91\,420 & O9\,V & 203309.45 & +411258.4 & 15.1 & 1.29 & 0.0 & $1.29^{+1.11}_{-.85}$ & $2.78 \times 10^{-6}$ & -- & -- & 0.01 & 0.05 & 0.17 & 2 \\
MT91\,421 & O9.5\,V & 203309.60 & +411300.6 & 126.1 & 1.29 & 0.0 & $0.57^{+.22}_{-.18}$ & $5.23 \times 10^{-5}$ & -- & -- & 0.06 & 0.80 & 0.43 & 22 \\
MT91\,448 & O6\,V & 203313.25 & +411328.6 & 187.7 & 1.41 & 0.0 & $0.31^{+.11}_{-.07}$ & $1.98 \times 10^{-4}$ & -- & -- & 0.06 & 3.82 & 1.01 & 36 \\
MT91\,455 & O8\,V & 203313.68 & +411305.7 & 173.3 & 1.21 & 0.0 & $0.70^{+.25}_{-.16}$ & $3.56 \times 10^{-5}$ & -- & -- & 0.07 & 0.65 & 0.81 & 29 \\
Cyg\,OB2 \#7 & O3\,If & 203314.11 & +412022.0 & 3550.2 & 1.00 & $0.81^{+.19}_{-.10}$ & $0.17^{+.02}_{-.03}$ & $3.74 \times 10^{-2}$ & $0.63^{+.18}_{-.05}$ & $8.31 \times 10^{-4}$ & 1.25 & 6.90 & 1.28 & 156\\
Cyg\,OB2 \#8b & O6.5\,III & 203314.76 & +411841.7 & 590.5 & 0.97 & $0.40^{+.16}_{-.25}$ & $0.31^{+.16}_{-.05}$ & $5.69 \times 10^{-4}$ & -- & -- & 0.18 & 1.89 & 1.18 & 79 \\
Cyg\,OB2 \#23 & O9.5\,V & 203315.77 & +412017.0 & 61.4 & 1.00 & 0.0 & $0.45^{+.21}_{-.17}$ & $1.61 \times 10^{-5}$ & -- & -- & 0.02 & 0.41 & 0.46 & 11 \\
Cyg\,OB2 \#8d & O8.5\,V & 203316.34 & +411902.0 & 149.6 & 0.99 & 0.0 & $0.61^{+.16}_{-.33}$ & $2.83 \times 10^{-5}$ & -- & -- & 0.07 & 0.84 & 0.93 & 30 \\
Cyg\,OB2 \#24 & O7.5\,V & 203317.48 & +411709.2 & 196.9 & 1.08 & 0.0 & $0.61^{+.13}_{-.13}$ & $3.24 \times 10^{-5}$ & -- & -- & 0.07 & 0.96 & 0.76 & 32 \\
Cyg\,OB2 \#8c & O5\,III & 203317.99 & +411831.2 & 981.5 & 0.96 & $0.43^{+.13}_{-.13}$ & $0.27^{+.08}_{-.08}$ & $1.29 \times 10^{-3}$ & $0.93^{+.66}_{-.21}$ & $5.37 \times 10^{-5}$ & 0.40 & 3.59 & 0.90 & 95 \\
MT91\,485 & O8\,V & 203318.02 & +412136.8 & 225.3 & 1.03 & $0.45^{+.50}_{-.45}$ & $0.12^{+.19}_{-.03}$ & $2.18 \times 10^{-2}$ & $0.98^{+1.13}_{-.26}$ & $1.61 \times 10^{-5}$ & 0.08 & 2.50 & 0.78 & 36 \\
MT91\,507 & O8.5\,V & 203321.01 & +411740.1 & 94.3 & 1.04 & 0.0 & $0.52^{+.32}_{-.20}$ & $1.63 \times 10^{-5}$ & -- & -- & 0.03 & 0.46 & 0.87 & 18 \\ 
MT91\,516 & O5.5\,V & 203323.48 & +410912.6 & 7349.5 & 1.44 & $0.19^{+.08}_{-.08}$ & $0.30^{+.20}_{-.05}$ & $3.64 \times 10^{-3}$ & $1.87^{+.16}_{-.09}$ & $7.25 \times 10^{-4}$ & 4.16 & 35.10 & 1.10 & 306 \\
Cyg\,OB2 \#25 & O8.5\,V & 203325.53 & +413326.6 & 65.4 & 1.06 & $0.52^{+.72}_{-.43}$ & $0.19^{+.15}_{-.07}$ & $7.88 \times 10^{-4}$ & -- & -- &  0.03 & 0.49 & 1.34 & 15 \\
MT91\,534 & O7.5\,V & 203326.74 & +411059.4 & 872.2 & 1.24 & $0.83^{+.37}_{-.33}$ & $0.16^{+.06}_{-.03}$ & $1.21 \times 10^{-2}$ & $2.25^{+.42}_{-.25}$ &  $9.52 \times 10^{-5}$ & 0.55 & 1.73 & 1.12 & 123 \\
Cyg\,OB2 \#74 & O8\,V & 203330.30 & +413557.9 & 127.0 & 1.26 & 0.0 & $0.60^{+.12}_{-.14}$ & $4.72 \times 10^{-5}$ & -- & -- & 0.08 & 1.40 & 0.77 & 24 \\
MT91\,611 & O7\,V & 203340.90 & +413017.9 & 21.9 & 1.06 & 0.0 & $0.79^{+.54}_{-.72}$ & $5.0 \times 10^{-6}$ & -- & -- & 0.02 & 0.15 & 1.22 & 2 \\
Cyg\,OB2 \#10 & O9\,I & 203346.11 & +413300.7 & 253.6 & 1.04 & 0.0 & $0.60^{+.13}_{-.09}$ & $5.95 \times 10^{-5}$ & -- & -- & 0.13 & 1.77 & 1.22 & 46 \\
Cyg\,OB2 \#27 &  O9.5\,V + BO\,V & 203359.56 & +411735.5 & 113.3 & 1.11 & 0.0 & $0.60^{+.15}_{-.15}$ & $2.38 \times 10^{-5}$ & -- & -- & 0.05 & 0.70 & 0.70 & 22 \\
Cyg\,OB2 \#41 & O9\,V & 203404.87 & +410513.1 & 31.7 & 1.37 & 0.0 & $0.93^{+.67}_{-.56}$ & $8.1 \times 10^{-6}$ & -- & -- & 0.02 & 0.23 & 0.97 & 5 \\
Cyg\,OB2 \#11 & O5If+B0V & 203408.52 & +413659.3 & 1262.9 & 1.03 & $0.64^{+.15}_{-.13}$ & $0.20^{+.04}_{-.04}$ & $1.02 \times 10^{-2}$ & $0.91^{+.11}_{-.14}$ & $3.22 \times 10^{-4}$ & 1.03 & 6.38 & 1.16 & 119 \\
Cyg\,OB2 \#75 & O9\,V & 203409.51 & +413413.9 & 185.9 & 1.00 & $0.50^{+.25}_{-.50}$ & $1.21^{+1.02}_{-.26}$ & $4.18 \times 10^{-5}$ & -- & -- & 0.14 & 0.35 & 1.12 & 33 \\
Cyg\,OB2 \#29 & O7\,V & 203413.53 & +413502.8 & 147.3 & 1.03 & 0.0 & $0.71^{+.16}_{-.23}$ &  $3.76 \times 10^{-5}$ & -- & -- & 0.11 & 1.15 & 0.59 & 23 \\
CP2012\,E54 & O9.5\,V & 203416.03 & +410219.5 & 46.8 & 1.22 & $1.16^{+2.17}_{-.96}$ & $0.19^{+.33}_{-.13}$ & $2.20 \times 10^{-3}$ & -- & -- & 0.02 & 0.16 & 1.75 & 8 \\
Cyg\,OB2 \#73 & O8\,III + O8\,III & 203421.95 & +411701.5 & 251.4 & 1.14 & $0.32^{+.17}_{-.15}$ & $0.92^{+.15}_{-.19}$ & $7.48 \times 10^{-5}$ & -- & -- &  0.20 & 0.92 & 1.10 & 41 \\
MT91\,771 & O7\,V + O9\,V & 203429.60 & +413145.3 & 1223.4 & 1.34 & $0.18^{+.14}_{-.11}$ & $0.55^{+.07}_{-.11}$ & $5.19 \times 10^{-4}$ & $1.82^{+.76}_{-.39}$ & $1.06 \times 10^{-4}$ & 1.06 & 8.95 & 0.95 & 128 \\
\hline
\end{tabular}
\end{sidewaystable}

\begin{sidewaystable}
\begin{center}
\caption{Same as Table\,\ref{fitsO}, but for B-type stars}
\tiny
\begin{tabular}{l c c c c c c c c c c c c}
\hline
\multicolumn{1}{c}{Star} & Spectral type & $\alpha$ & $\delta$ & Total counts & $N_{\rm H}^{ISM}$ & $N_{\rm wind}$ & $kT_1$ & norm$_1$ & $f_{\rm X}^{\rm obs}$ & $f_{\rm X}$ & $\chi^2_{\nu}$ & $\nu$\\
& & (J2000.0) & (J2000.0) & & ($10^{22}$\,cm$^{-2}$) & ($10^{22}$\,cm$^{-2}$) & (keV) & & \multicolumn{2}{c}{($10^{-14}$\,erg\,cm$^{-2}$\,s$^{-1}$)} & & \\
\hline
MT91\,20 & B0\,V + O9\,I & 203051.07 & +412021.6 & 10.7 & 1.37 & -- & -- & -- & 0.16 & 0.36 & -- & -- \\
MT91\,42 & B2\,V & 203059.56 & +413600.2 & 5.8 & 0.84 & -- & -- & -- & 0.26 & 0.50 & -- & -- \\
CPR\,2002 A30 & B2\,V & 203122.11 & +411202.9 & 4.0 & 1.09 & -- & -- & -- & 0.62 & 1.31 & -- & -- \\
MT91\,103 & B1\,V + B2\,V & 203133.35 & +412248.2 & 144.8 & 1.26 & 0.0 & $2.46^{+.86}_{-.52}$ & $2.10 \times 10^{-5}$ & 1.42 & 3.10 & 1.14 & 30 \\
MT91\,129 & B3\,V & 203141.66 & +412820.3 & 49.3 & 0.86 & 0.0 & $3.0^{+11.3}_{-1.4}$ & $6.2 \times 10^{-6}$ & 0.55 & 0.96 & 0.84 & 7 \\
MT91\,174 & B2\,III & 203156.96 & +413148.0 &   6.8 & 0.80 & -- & -- & -- & 0.05 & 0.10 & -- & --\\
MT91\,179 & B3\,V & 203159.93 & +413712.8 & 4.4 & 0.79 & -- & -- & -- & 0.07 & 0.14 & -- & -- \\
MT91\,213 & B0\,V   & 203213.13 & +412724.3 & 270.6 & 0.79   & 0.0 & $3.97^{+2.06}_{-.96}$ & $2.4 \times 10^{-5}$ & 2.62 & 4.05 & 1.11 & 48\\ 
MT91\,216 & B1.5\,V & 203213.82 & +412741.6 & 31.1 & 0.80 & 0.0 & $2.63^{+4.38}_{-1.91}$ & $3.9 \times 10^{-6}$ & 0.32 & 0.58 & 0.48 & 5 \\
MT91\,220 & B1\,V & 203214.61 & +412233.5 & 8.0 & 1.00 & -- & -- & -- & 0.08 & 0.17 & -- & -- \\
MT91\,221 & B2\,V & 203214.70 & +412739.6 & 203.3 & 0.86 & $0.51^{+.34}_{-.50}$ & $1.21^{+.64}_{-.44}$ & $2.95 \times 10^{-5}$ & 1.09 & 2.40 & 1.00 & 36 \\
MT91\,239 & B4\,V & 203221.77 & +413425.4 & 7.8 & 0.74 & -- & -- & -- & 0.09 & 0.16 & -- & -- \\
MT91\,250 & B2\,III & 203226.10 & +412940.9 &   7.4 & 0.72 & -- & -- & -- & 0.06 & 0.11 & -- & --\\
MT91\,252 & B1.5\,III + B1\,V & 203226.53 & +411913.4 & 14.4 & 1.00 & -- & -- & -- & 0.21 & 0.42 & -- & -- \\
MT91\,255 & B2\,III & 203227.25 & +412156.6 & 35.7 & 0.92 & 0.0 & $3.0^{+11.4}_{-2.0}$ & $4.8 \times 10^{-6}$ & 0.41 & 0.74 & 0.47 & 9 \\
Cyg\,OB2 \#21 & B0.5\,V & 203227.77 & +412852.1 & 46.2 & 0.71 & 0.0 & $0.61^{+.26}_{-.35}$ & $6.0 \times 10^{-6}$ & 0.23 & 1.79 & 0.74 & 8 \\
MT91\,271 & B4\,V & 203232.42 & +412257.9 & 7.1 & 0.95 & -- & -- & -- & -- & -- & -- & -- \\
MT91\,295 & B2\,V & 203237.72 & +412615.5 & 43.3 & 0.81 & 0.0 & $4.4^{+...}_{-2.2}$ & $3.8 \times 10^{-6}$ & 0.44 & 0.67 & 0.65 & 10 \\
MT91\,298 & B3\,V & 203238.35 & +412857.0 & 43.7 & 0.81 & $0.21^{+1.07}_{-.21}$ & $1.60^{+1.09}_{-.72}$ & $6.7 \times 10^{-6}$ & 0.37 & 0.75 & 0.59 & 7 \\
MT91\,300 & B1\,V & 203238.89 & +412520.3 & 22.1 & 0.82 & 0.0 & $4.35^{+...}_{-3.90}$ & $2.8 \times 10^{-6}$ & 0.31 & 0.48 & 0.67 & 3\\
CPR\,2002 A31 & B0.5\,V & 203239.50 & +405247.5 & 18.6 & 1.29 & 0.0 & $0.27^{+.99}_{-.27}$ & $4.0 \times 10^{-5}$ & 0.09 & 7.19 & 1.31 & 21 \\ 
MT91\,311 & B2\,V + B3\,V & 203242.88 & +412016.4 & 194.2 & 0.90 & $0.39^{+.23}_{-.26}$ & $1.10^{+.23}_{-.16}$ & $2.1 \times 10^{-5}$ & 0.74 & 2.10 & 1.15 & 36 \\
MT91\,322 & B2.5\,V & 203246.47 & +412422.0 & 93.6 & 0.87 & $0.55^{+.35}_{-.39}$ & $1.15^{+.48}_{-.23}$ & $2.36  \times 10^{-5}$ & 0.80 & 1.83 & 0.67 & 17 \\
MT91\,336 & B3\,III & 203249.65 & +412536.4 & 97.6 & 0.77 & 0.0 & $2.39^{+1.16}_{-.86}$ & $1.37 \times 10^{-5}$ & 1.07 & 2.01 & 0.83 & 17 \\
Cyg\,OB2 \#37 & B3\,V & 203254.40 & +411521.9 & 14.5 & 1.33 & -- & -- & -- & 0.10 & 0.23 & -- & -- \\
MT91\,372 & B0\,V + B2\,V & 203258.90 & +410430.1 & 72.4 & 1.37 & 0.0 & $1.97^{+1.91}_{-.82}$ & $1.68 \times 10^{-5}$ & 0.91 & 2.47 & 0.98 & 13 \\
MT91\,378 & B0\,V & 203259.63 & +411514.7 & 61.9 & 1.33 & 1.44 & $0.18^{+.12}_{-.13}$ & $2.43 \times 10^{-3}$ & 0.13 & 0.79 & 0.93 & 15 \\
MT91\,400 & B1\,V & 203305.16 & +411751.2 & 7.4 & 1.05 & -- & -- & -- & 0.04 & 0.09 & -- & -- \\ 
MT91\,425 & B0\,V & 203310.10 & +411310.2 & 17.0 & 1.24 & $2.20_{-2.20}^{+6.95}$ & $0.07^{+2.08}_{-.04}$ & 29.7 & 0.03 & 0.41 & 0.32 & 3 \\ 
MT91\,428 & B1\,V & 203310.47 & +412057.4 & 32.8 & 1.14 & 0.0 & $3.6^{+...}_{-2.8}$ & $3.3 \times 10^{-6}$ & 0.31 & 0.54 & 1.06 & 6 \\
MT91\,429 & B0\,V + B3\,V & 203310.57 & +412222.7 & 11.7 & 1.02 & -- & -- & -- & 0.08 & 0.13 & -- & -- \\
MT91\,435 & B0\,V & 203311.09 & +411032.3 & 14.2 & 1.38 & $0.11^{+3.68}_{-0.10}$ & $0.42^{+0.97}_{-.32}$ & $9.0 \times 10^{-6}$ & 0.05 & 1.37 & 0.42 & 3 \\ 
MT91\,453 & B5\,V & 203313.34 & +412639.4 & 15.5 & 0.73 & 0.0 & $2.32^{+...}_{-1.95}$ & $2.1 \times 10^{-6}$ & 0.17 & 0.31 & 0.59 & 2 \\ 
MT91\,459 & B5\,V & 203314.33 & +411933.0 & 104.5 & 1.10 & $0.42^{+.46}_{-.42}$ & $0.30^{+.26}_{-.11}$ & $1.26 \times 10^{-4}$ & 0.28 & 3.69 & 1.11 & 20 \\
MT91\,467 & B1\,V & 203315.27 & +412956.5 & 52.4 & 1.03 & 0.0 & $3.16^{+4.91}_{-1.46}$ & $9.0 \times 10^{-6}$ & 0.79 & 1.42 & 0.65 & 9 \\
MT91\,477 & B0\,V & 203317.41 & +411238.7 & 17.9 & 1.23 & $1.23^{+4.88}_{-1.23}$ & $0.38^{+1.45}_{-.32}$ & $3.37 \times 10^{-5}$ & 0.06 & 0.23 & 0.65 & 3 \\
MT91\,509 & B0\,III & 203321.02 & +413552.4 & 0.8 & 1.15 & -- & -- & -- & 0.04 & 0.10 & -- & -- \\
Cyg\,OB2 \#18 & B1\,Ib & 203330.77 & +411522.7 & 28.3 & 1.10 & 0.0 & $0.18^{+.69}_{-.09}$ & $9.1 \times 10^{-5}$ & 0.06 & 10.70 & 1.06 & 9 \\
MT91\,561 & B2\,V & 203331.62 & +412146.7 & 6.6 & 0.86 & -- & -- & -- & -- & -- & -- & -- \\ 
MT91\,573 & B3\,I & 203333.97 & +411938.1 & 21.2 & 0.98 & 0.0 & $4.02^{+...}_{-3.29}$ & $5.4 \times 10^{-6}$ & 0.57 & 0.92 & 0.95 & 2 \\
Cyg\,OB2 \#70 & B0\,V & 203337.00 & +411611.1 & 91.5 & 1.09 & $0.13^{+.42}_{-.13}$ & $0.59^{+.25}_{-.23}$ & $1.90 \times 10^{-5}$ & 0.31 & 3.47 & 0.85 & 24 \\
CP2012\,E52 & B0\,Ib & 203338.21 & +405341.1 & 101.4 & 1.10 & 0.0 & $0.97^{+.21}_{-.25}$ & $3.44 \times 10^{-5}$ & 1.29 & 9.24 & 0.64 & 17 \\
Cyg\,OB2 \#19 & B0\,Iab & 203339.09 & +411925.9 & 159.2 & 0.97 & $0.28^{+.34}_{-.22}$ & $0.41^{+.14}_{-.14}$ & $7.52 \times 10^{-5}$ & 0.59 & 5.96 & 0.71 & 27 \\ 
MT91\,620 & B0\,V & 203342.30 & +411146.4 & 31.0 & 1.17 & 0.0 & $0.81^{+.52}_{-.54}$ & $3.1 \times 10^{-6}$ & 0.09 & 0.92 & 1.71 & 6 \\
MT91\,621 & B1\,V & 203342.55 & +411457.0 & 9.1 & 1.21 & -- & -- & -- & 0.10 & 0.22 & -- & -- \\
MT91\,635 & B1\,III & 203346.83 & +410801.6 & 169.1 & 1.10 & 0.0 & $2.58^{+.96}_{-1.00}$ & $1.35 \times 10^{-5}$ & 0.99 & 2.01 & 1.23 & 38 \\
Cyg\,OB2 \#26 & B1\,III & 203347.81 & +412041.2 & 73.2 & 1.01 & $0.86^{+.65}_{-.40}$ & $0.73^{+.31}_{-.24}$ & $2.41 \times 10^{-5}$ & 0.32 & 0.89 & 0.87 & 17 \\
MT91\,639 & B2\,V & 203347.84 & +410908.2 & 30.3 & 1.12 & 0.0 & $1.1^{+16.1}_{-.8}$ & $3.1 \times 10^{-6}$ & 0.13 & 0.70 & 0.93 & 7 \\
MT91\,646 & B1.5\,V & 203348.83 & +411940.5 & 107.9 & 0.92 & $1.42^{+1.37}_{-.73}$ & $1.64^{+.69}_{-.56}$ & $2.22 \times 10^{-5}$ & 0.77 & 1.03 & 0.88 & 22 \\
MT91\,692 & B0\,V & 203359.23 & +410537.9 & 6.7 & 1.10 & -- & -- & -- & 0.26 & 0.54 & -- & -- \\
MT91\,720 & B0.5\,V + B1.5\,V & 203406.02 & +410809.3 & 7.2 & 1.29 & -- & -- & -- & 0.09 & 0.21 & -- & -- \\
MT91\,759 & B1\,V & 203424.60 & +412624.7 & 142.6 & 1.08 & 0.0 & $2.59^{+1.13}_{-.63}$ & $2.33 \times 10^{-5}$ & 1.74 & 3.48 & 0.82 & 24 \\
CPR\,2002 A36 & B0\,Ib + B0\,III & 203458.78 & +413617.3 & 269.3 & 1.19 & 0.0 & $0.90^{+.10}_{-.22}$ & $1.25 \times 10^{-4}$ & 4.01 & 36.01 & 0.64 & 40 \\
\hline
\end{tabular}
\end{center}
\end{sidewaystable}

\begin{sidewaystable}
\caption{Same as Table\,\ref{fitsO}, but for Wolf-Rayet stars}
\tiny
\begin{tabular}{l c c c c c c c c c c c c c c c}
\hline
\multicolumn{1}{c}{Star} & Spectral type & $\alpha$ & $\delta$ & Total counts & $N_{\rm H}^{ISM}$ & Abund & $N_{\rm wind}$ & $kT_1$ & norm$_1$ & kT$_2$ & norm$_2$ & $f_{\rm X}^{\rm obs}$ & $f_{\rm X}$ & $\chi^2_{\nu}$ & $\nu$ \\
& & (J2000.0) & (J2000.0) & & ($10^{22}$\,cm$^{-2}$) & & ($10^{22}$\,cm$^{-2}$) & (keV) & & (keV) & & \multicolumn{2}{c}{($10^{-13}$\,erg\,cm$^{-2}$\,s$^{-1}$)} & & \\
\hline
WR\,144 & WC4 & 203202.92 & +411518.8 & 5.8 & 0.97 & -- & -- & -- & -- & -- & -- & -- & -- & -- & --\\
WR\,145 & WN7o/CE + O7\,V((f)) & 203206.26 & +404829.6 & 696.3 & 0.97 & WC & $(3.4_{-.9}^{+.6}) \times 10^{-3}$ & $1.59^{+.38}_{.17}$ & $4.05 \times 10^{-6}$ & -- & -- & 5.39 & 6.56 & 1.08 & 110 \\
WR\,145 & WN7o/CE + O7\,V((f)) & 203206.26 & +404829.6 & 696.3 & 0.97 & solar & $2.68_{-.73}^{+.53}$ & $1.60^{+.43}_{.19}$ & $2.12 \times 10^{-3}$ & -- & -- & 5.33 & 6.50 & 1.05 & 110 \\
WR\,146 & WC6 + O8\,III & 203547.07 & +412244.7 & 1827.5 & 1.32 & WC & $(6.2^{+2.1}_{-1.7}) \times 10^{-4}$ & $0.36^{+.09}_{.08}$ & $4.43 \times 10^{-6}$ & $2.10^{+.25}_{-.22}$ & $8.28 \times 10^{-7}$ & 2.92 & 13.02 & 1.32 & 176 \\ 
WR\,146 & WC6 + O8\,III & 203547.07 & +412244.7 & 1827.5 & 1.32 & solar & $0.46^{+.20}_{-0.17}$ & $0.36^{+.11}_{.08}$ & $2.88 \times 10^{-3}$ & $2.14^{+.40}_{-.27}$ & $4.06 \times 10^{-4}$ & 2.95 & 13.60 & 1.23 & 176 \\ 
\hline
\end{tabular}
\end{sidewaystable}

\end{document}